\documentclass[aps,prd,floats,floatfix,twoside,preprintnumbers,superscriptaddress,nofootinbib]{revtex4}

\usepackage{graphicx} 
\usepackage{dcolumn}
\usepackage{amssymb}
\usepackage{amsmath}
\usepackage[hidelinks]{hyperref}
\usepackage{color}
\usepackage[utf8]{inputenc}
\usepackage{slashed}

\newcommand{\be}{\begin{equation}}
\newcommand{\ee}{\end{equation}}

\newcommand{\beq}{\begin{eqnarray}}
\newcommand{\eeq}{\end{eqnarray}}

\newcommand{\calG}{\mathcal{G}}
\newcommand{\calA}{\mathcal{A}}
\newcommand{\calF}{\mathcal{F}}

\newcommand{\calV}{\mathcal{V}}
\newcommand{\calS}{\mathcal{S}}
\newcommand{\ltwop}{\boldsymbol{l}_{2T}}

\newcommand{\hatx}{\hat{x}} 
\newcommand{\hatz}{\hat{z}} 

\newcommand{\dhsig}{\Delta\hat{\sigma}}

\newcommand{\pd}{\partial}

\begin{document}
\date{\today}
\preprint{ZTF-EP-21-05}

\title{The $g_T(x)$ contribution to  single spin asymmetries in SIDIS}

\author{Sanjin Beni\' c}
\affiliation{Department of Physics, Faculty of Science, 
University of Zagreb, Bijeni\v cka c. 32, 10000 Zagreb, Croatia}

\author{Yoshitaka Hatta}
\affiliation{Physics Department, Brookhaven National Laboratory, Upton, New York 11973, USA}
\affiliation{RIKEN BNL Research Center, Brookhaven National Laboratory, Upton, New York, 11973, USA}

\author{Abhiram Kaushik}
\affiliation{Department of Physics, Faculty of Science, 
University of Zagreb, Bijeni\v cka c. 32, 10000 Zagreb, Croatia}

\author{Hsiang-nan Li}
\affiliation{Institute of Physics, Academia Sinica,
Taipei, Taiwan 11529, Republic of China}

\begin{abstract} 
Motivated by a novel origin of transverse single spin asymmetry (SSA) in semi-inclusive Deep Inelastic Scattering (SIDIS) uncovered by some of us, we quantitatively investigate its impact on the theoretical understanding of the mechanism responsible for SSA. This new contribution from the quark-initiated channel first appears in two-loop perturbation theory and involves the $g_T(x)$ distribution. We point out another entirely analogous piece from the gluon-initiated channel proportional to the gluon helicity distribution $\Delta G(x)$. Both contributions are solely expressed in terms of twist-two polarized parton distribution functions and twist-two fragmentation functions in the Wandzura-Wilczek approximation, such that they can be unambiguously evaluated without introducing free parameters. We make predictions for measurements of the asymmetries $A_{UT}$ at the future Electron-Ion Collider (EIC), and find that $A_{UT}$ associated with the $\sin (\phi_h-\phi_S)$, $\sin \phi_S$ and $\sin (2\phi_h-\phi_S)$ harmonics can reach up to 1-2\% even at the top EIC energy. 

    
\end{abstract}


\maketitle 

\section{Introduction}

Recently, three of us, together with D.~J.~Yang,  have proposed a novel mechanism for generating transverse single-spin asymmetry (SSA) in 
semi-inclusive deep inelastic scattering (SIDIS) $ep^\uparrow\to e'hX$ \cite{Benic:2019zvg}.  It has been demonstrated that an imaginary phase 
necessary for SSA can be produced purely within a parton-level cross section starting at two loops. The spin-dependent part of the cross section at high transverse momentum $P_{hT}>1$ GeV (measured with respect to the virtual photon direction) can be schematically written as 
\be
\frac{d\Delta \sigma}{dP_{hT}} \sim g_T(x)\otimes H\otimes D_1(z)+\cdots, \label{sche}
\ee
where $g_T(x)$ is the twist-three parton distribution function (PDF) associated with a transversely polarized proton, $D_1$ is the unpolarized twist-two fragmentation function (FF) for the observed hadron $h$, and $H$ is the hard kernel starting at ${\cal O}(\alpha_s^2)$ (see also an earlier related work \cite{Ratcliffe:1985mp}). The terms omitted in (\ref{sche}) are proportional to the `genuine twist-three' quark-gluon correlation functions $\sim \langle \bar{\psi}gF\psi\rangle$ commonly called the Efremov-Teryaev-Qiu-Sterman (ETQS) functions \cite{Efremov:1981sh,Qiu:1998ia}.  
As is well known, the $g_T$ distribution can be written as the sum of the Wandzura-Wilczek (WW) part \cite{Wandzura:1977qf} and the genuine twist-three part
\be
g_T(x) = \int_x^1 \frac{dx'}{x'} \Delta q(x') + ({\rm genuine\ twist\ three})\,,  \label{ww}
\ee
where $\Delta q(x)$ is the twist-two polarized (helicity) quark PDF. It is a consistent truncation of the result in Ref.~\cite{Benic:2019zvg} to keep only the WW part in (\ref{ww}).  The new source of SSA can then be entirely expressed in terms of the twist-two PDFs $\Delta q(x)$ and the twist-two FFs $D_1(z)$. This is a remarkable observation in striking contrast to the prevailing view that SSA at high-$P_{hT}$ is explained by the ETQS functions and certain twist-three fragmentation functions (see a review \cite{Pitonyak:2016hqh} and references therein).  Unlike these higher-twist distributions, twist-two distributions are very well constrained by global QCD analyses.  Therefore, the mechanism proposed in \cite{Benic:2019zvg} offers a unique part of SSA that can be unambiguously calculated without introducing free parameters.   Moreover, in the transverse-momentum-dependent (TMD) PDF framework valid in the low-$P_{hT}$ region ($P_{hT}\lesssim 1$ GeV),  a new source of SSA proportional to the $g_{1T}(x,k_\perp)$ distribution (the TMD version of $g_T(x)$) has been identified, along with more than a dozen of new contributions involving various twist-three TMDs and FFs and hard kernels up to two loops. Again, this calls into question  the prevailing view in the community (see, e.g., \cite{Cammarota:2020qcw}) that SSA at low-$P_{hT}$ is entirely attributed to the Sivers and Collins functions.     

The purpose of this paper is twofold. First, we extend the analysis of \cite{Benic:2019zvg} to gluon-initiated channels.   There exists a gluonic counterpart of $g_T(x)$,  the twist-three  ${\cal G}_{3T}(x)$ distribution \cite{Ji:1992eu,Hatta:2012jm} for a transversely polarized proton. Its WW part is related to  the twist-two polarized gluon PDF $\Delta G(x)$. In complete analogy to  (\ref{sche}), we find the structure 
\be
\frac{d\Delta \sigma}{dP_{h\perp}} \sim  {\cal G}_{3T}(x)\otimes H_g\otimes D_1(z) \sim \Delta G(x)\otimes H_g\otimes D_1(z)\,,  \label{newglue}
\ee 
 which again consists  only of twist-two distributions after the WW approximation. We shall identify the two-loop diagrams that go into the hard kernel $H_g$ and study their gauge invariance and infrared safety. Equation~\eqref{newglue} is a novel  gluon-initiated source of SSA in SIDIS to be considered jointly with the previously known mechanism which involves genuine twist-three, three-gluon correlators $\langle FgFF\rangle$ \cite{Kang:2008qh,Beppu:2010qn,Koike:2011ns}. 

Second, we perform a detailed numerical analysis of SSA and make predictions for its measurements at the future Electron-Ion Collider (EIC) \cite{Proceedings:2020eah,AbdulKhalek:2021gbh}. In doing so, we neglect the `usual' contributions from the ETQS and twist-three FFs, which have been intensively anallyzed, and focus on the new contributions in order to explore their importance clearly. The results can be viewed as a baseline for future EIC measurements of SSA. Deviations from our predictions, if observed, may be attributed to genuinely twist-three effects.   

This paper is organized as follows. In Section II, we describe the SIDIS setup and introduce kinematic variables. In Section III, we first review the result of \cite{Benic:2019zvg} obtained for the quark-initiated  channel, and then propose an analogous, but novel contribution to SSA in the gluon-initiated channel. 
In Section IV, we perform a detailed analysis of the two-loop diagrams and calculate the hard coefficients in all the partonic channels. In Section V, we implement the obtained formulas numerically and make predictions according to the kinematic coverage of the EIC. We also present results relevant to the COMPASS experiments \cite{Adolph:2014zba}. Finally, we discuss our findings and conclude in Section VI. Appendices are devoted to a technical proof of the infrared finiteness of the factorization formulas at two-loop level.

\section{SIDIS Kinematics}
\label{sec:kin}

In this section we give a brief review of polarized SIDIS $e(l)p(P)\to e(l')h(P_h)X$ and introduce involved kinematic variables. We have in mind light hadron production specifically for $h=\pi^\pm$. Heavy-quark production will be studied in a separate work. 
The spin-dependent part of the differential cross section is given by
\be
d^6\Delta\sigma = \frac{1}{2S_{ep}}\frac{d^3P_h}{(2\pi)^32E_h}\frac{d^3 l'}{(2\pi)^32E_{l'}} \frac{e^4}{(Q^2)^2} L^{\mu\nu}W_{\mu\nu}\,,
\label{eq:xspol}
\ee
where $S_{ep}\equiv (l+P)^2$, $Q^2\equiv -q^2=-(l-l')^2$, $L^{\mu\nu}=2(l^\mu l'^\nu + l^\nu l'^\mu)-g^{\mu\nu}Q^2$ is the leptonic tensor, $W^{\mu\nu}$ is the hadronic tensor, and $\nu$ and $\mu$ are  the polarization indices of the virtual photon in the amplitude and the complex-conjugate amplitude, respectively. The Bjorken variable is denoted as $x_B=Q^2/(2P\cdot q)$.  We shall work in the so-called hadron frame, where the virtual photon and the proton move in the $z$ direction with
\be
q^\mu=(0,0,0,-Q)\,, \qquad P^\mu=\left(\frac{Q}{2x_B},0,0, \frac{Q}{2x_B} \right)\,.
\ee
The incoming and outgoing leptons have the momenta
\be
l^\mu = \frac{Q}{2} (\cosh\psi ,\sinh\psi \cos \phi, \sinh\psi \sin \phi, -1)\,, \qquad l'^\mu = \frac{Q}{2} (\cosh\psi ,\sinh\psi \cos \phi, \sinh\psi \sin \phi, 1)\,,
\label{eq:use}
\ee
where $\phi$ is the azimuthal angle relative to the $z$ axis, and
\be
\cosh\psi \equiv \frac{2x_BS_{ep}}{Q^2}-1\,.
\ee
We adopt the standard variables 
\be
y=\frac{P\cdot q}{P\cdot l}\,, \qquad  z_f =\frac{ P\cdot P_h}{ P \cdot q}\,,  
\ee
with the relation $x_ByS_{ep}=Q^2$. Another common variable is $q_T = \sqrt{-q_t^2}$ where
\be
q^\mu_t \equiv q^\mu - \frac{P_h \cdot q}{P_h\cdot P} P^\mu - \frac{P \cdot q}{P \cdot P_h} P_h^\mu \,.
\ee
In the present frame, the transverse part of $q^\mu_t$ reads $\vec{q}_{tT} = -\vec{P}_{hT}/z_f$. The momentum of the final state hadron can then be parametrized as
\be
P_h^\mu = \frac{z_f Q}{2}\left(1 + \frac{q_T^2}{Q^2}, \frac{2 q_T}{Q} \cos\chi , \frac{2 q_T}{Q} \sin \chi , - 1 + \frac{q_T^2}{Q^2}\right)\,.
\label{eq:Ph}
\ee
For the transverse spin of the incoming proton we choose 
\be
S_T^\mu = (0,\cos\Phi_S,\sin\Phi_S,0)\,. 
\ee

In terms of the above variables,  the differential cross section \eqref{eq:xspol} takes the following Lorentz invariant form  
\be
\frac{d^6\Delta \sigma}{dx_B dQ^2 d z_f d q_T^2 d\phi d \chi} = \frac{\alpha_{\rm em}^2}{128 \pi^4 x_B^2 S_{ep}^2 Q^2} z_f L_{\mu\nu} W^{\mu\nu}\,.
\label{eq:xspol2}
\ee
In practice, instead of $\phi$ and $\chi$, it is more convenient to define the  hadron and spin angles relative to the lepton plane,
\beq
 \phi_h\equiv \phi-\chi\,, \qquad \phi_S\equiv \phi-\Phi_S\,,
\eeq
in accordance with the Trento conventions \cite{Bacchetta:2004jz}. The cross section is then a function of $\phi_h$ and $\Phi_S-\chi = \phi_h-\phi_S$. The dependence on $\phi_h$ can be factored out by 
decomposing  the hadron tensor $W^{\mu\nu}$  using the following set of vectors \cite{Meng:1991da},
\be
\begin{split}
& T^\mu = \frac{1}{Q}\left(q^\mu + 2 x_B P^\mu\right)\,,\\
& X^\mu = \frac{1}{q_T}\left[\frac{P_h^\mu}{z_f} - q^\mu - \left(1 + \frac{q_T^2}{Q^2}\right) x_B P^\mu\right]\,,\\
& Y^\mu = \epsilon^{\mu\nu\rho\sigma} Z_\nu X_\rho T_\sigma\,,\\
& Z^\mu = - \frac{q^\mu}{Q}\,,
\end{split}
\ee
which form nine independent tensors, $\calV_k^{\mu\nu}$ (see \cite{Meng:1991da} for explicit expressions), and their inverses, $\tilde{\calV}_k^{\mu\nu}$. Among them,  the following six symmetric tensors \cite{Meng:1991da} contribute to  the decomposition of $W^{\mu\nu}$,
\be
\begin{split}
& \tilde{\calV}_1^{\mu\nu} = \frac{1}{2}\left(2 T^\mu T^\nu + X^\mu X^\nu + Y^\mu Y^\nu\right)\,,\\
&\tilde{\calV}_2^{\mu\nu} =  T^\mu T^\nu\,,\\
&\tilde{\calV}_3^{\mu\nu} =  -\frac{1}{2}(T^\mu X^\nu + X^\mu T^\nu)\,,\\
&\tilde{\calV}_4^{\mu\nu} =  \frac{1}{2}(X^\mu X^\nu - Y^\mu Y^\nu)\,,\\
&\tilde{\calV}_8^{\mu\nu} =  -\frac{1}{2}(T^\mu Y^\nu + Y^\mu T^\nu)\,,\\
&\tilde{\calV}_9^{\mu\nu} =  \frac{1}{2}(X^\mu Y^\nu + Y^\mu X^\nu)\,.\\
\end{split}
\label{eq:basis}
\ee
With these tensors we can write
\be
L_{\mu\nu} W^{\mu\nu} = Q^2 \sum_{k = 1,\dots,4, 8,9} \calA_k(\phi_h) [W_{\rho \sigma} \tilde{\calV}_k^{\rho \sigma} ]\,,
\label{eq:lw}
\ee
where 
\be
\calA_k(\phi_h) = L_{\mu\nu} \calV_k^{\mu\nu}/Q^2\,,
\ee
have the explicit expressions
\be
\begin{split}
&\calA_1(\phi_h) = 1 + \cosh^2\psi  \,,\\
&\calA_2(\phi_h) = -2 \,,\\
&\calA_3(\phi_h) = - \cos\phi_h \sinh 2\psi \,,\\ 
&\calA_4(\phi_h) = \cos 2\phi_h \sinh^2 \psi \,,\\ 
&\calA_8(\phi_h) = - \sin\phi_h \sinh 2\psi\,,\\ 
&\calA_9(\phi_h) = \sin 2\phi_h \sinh^2 \psi\,.
\end{split}
\label{eq:Ak}
\ee

We are thus led to the representation (see for example \cite{Kanazawa:2013uia}) 
\be
\begin{split}
\frac{d^6\Delta \sigma}{dx_B dQ^2 d z_f d q_T^2 d\phi d \chi} & = \sin(\phi_h - \phi_S)\left(\calF_1 + \calF_2\cos\phi_h + \calF_3 \cos 2\phi_h\right) + \cos(\phi_h - \phi_S)\left(\calF_4\sin\phi_h + \calF_5 \sin 2\phi_h\right)\\
& =\Big[F^{\sin(\phi_h - \phi_S)} \sin(\phi_h - \phi_S) + F^{\sin(2\phi_h - \phi_S)}\sin(2\phi_h - \phi_S) + F^{\sin\phi_S} \sin \phi_S\\
& + F^{\sin(3\phi_h - \phi_S)} \sin(3\phi_h - \phi_S) + F^{\sin(\phi_h + \phi_S)} \sin(\phi_h + \phi_S)\Big]\,,\\
\end{split} \label{fourier}
\ee
with
\be
\begin{split}
& F^{\sin(\phi_h - \phi_S)} = \calF_1\,,\\
& F^{\sin(2\phi_h - \phi_S)} = \frac{\calF_2 + \calF_4}{2}\,,\\
& F^{\sin \phi_S} = \frac{-\calF_2 + \calF_4}{2}\,,\\
& F^{\sin(3\phi_h - \phi_S)} = \frac{\calF_3 + \calF_5}{2}\,,\\
& F^{\sin(\phi_h + \phi_S)} = \frac{-\calF_3 + \calF_5}{2}\,.\\
\end{split} \label{FF}
\ee
The Fourier components $\sin (\phi_h-\phi_S)$ and $\sin (\phi_h+\phi_S)$ are referred to as the Sivers and Collins asymmetries, respectively. While we continue to use this nomenclatures, we emphasize that the  new mechanism, which contributes to these asymmetries and will be studied in detail below, has nothing to do with the Sivers and Collins functions, or their collinear twist-three counterparts.

\section{New contributions to SSA}

In this section we first recapitulate the $g_T(x)$ contribution to SSA discussed in \cite{Benic:2019zvg}, and  apply the so-called Wandzura-Wilczek (WW) approximation to simplify the result. We then derive another new contribution to SSA due to the gluonic counterpart of $g_T$.

\subsection{Quark-initiated channel}

\begin{figure}
  \begin{center}
  \includegraphics[scale = 0.5]{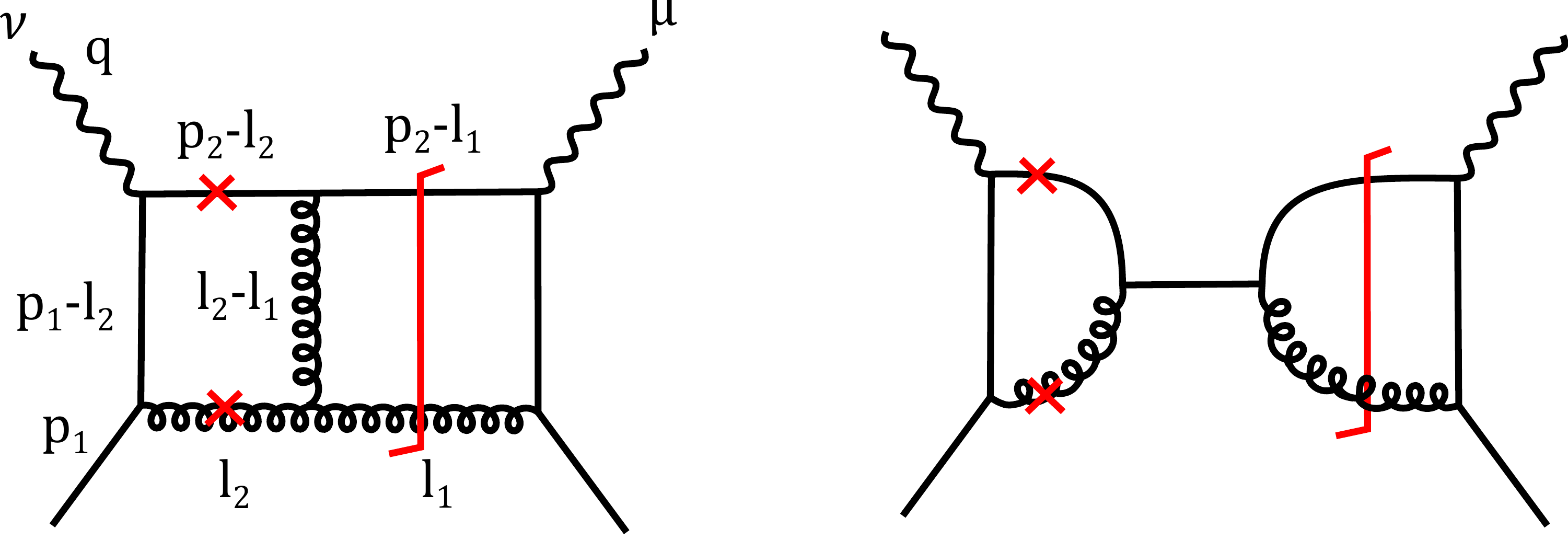}
  \end{center}
  \caption{Prototype two-loop diagrams contributing to SSA in the quark initiated channel, where the crosses denote the cuts needed to generate an imaginary phase, the vertical line is the final state cut, and $p_1=xP$ is the incoming quark momentum.}
  \label{fig:quark}
\end{figure}

In \cite{Benic:2019zvg}, it has been shown that the imaginary phase necessary for SSA in SIDIS can come from the hard kernel in perturbation theory starting at two loops, and all the relevant two-loop diagrams have been identified. However, only the quark (and antiquark) initiated  channel was  considered there.  In this channel,  motivated by the structure (\ref{sche}), we factorize the fragmentation function out of the hadronic tensor $W_{\mu\nu}$ as  
\be
W_{\mu\nu} = \sum_{a = q,\bar{q},g}\int \frac{dz}{z^2} D_1^a(z) w^a_{\mu\nu}\,,
\label{eq:WDfact}
\ee 
where we have taken into account the fact that the observed hadron can also come from the fragmentation of a radiated gluon in the final state.   
 The result of \cite{Benic:2019zvg} reads (suppressing the label $a$ for simplicity)
\be
\begin{split}
w_{\mu\nu} & = \frac{M_N}{2}\int dx g_T(x) {\rm Tr} \left[\gamma_5 \slashed{S}_T S^{(0)}_{\mu\nu}(x P) \right]\\
&- \frac{M_N}{4}\int dx \tilde{g}(x) {\rm Tr}\left[ \gamma_5\slashed{P} S_T^\alpha  \left.\frac{\partial S^{(0)}_{\mu\nu}(k)}{\partial k_T^\alpha}\right|_{k=xP} \right]\\
&+ \frac{iM_N}{4}\int dx_1 dx_2 {\rm Tr}\left[\left(\slashed{P} \epsilon^{\alpha P n S_T}  \frac{G_F(x_1,x_2)}{x_1-x_2} + i\gamma_5\slashed{P} S_T^\alpha  \frac{\tilde{G}_F(x_1,x_2)}{x_1-x_2} \right)  S_{\mu\nu\alpha}^{(1)}(x_1 P,x_2 P) \right]\,,
\end{split}
\label{eq:wtot}
\ee
in which $M_N$ is the proton mass, and ${\rm Tr}$ denotes trace over colors and Dirac indices. Our conventions are $\epsilon_{0123}=+1$, $\gamma_5=i\gamma^0\gamma^1\gamma^2\gamma^3$ and $\epsilon^{\alpha PnS} =\epsilon^{\alpha\beta\rho\lambda}P_\beta n_\rho S_{T\lambda}$ with the light-like vector $n^\mu$ satisfying  $n^2 = 1$ and $n\cdot P = 1$.  
The $g_T(x)$ distribution function is defined as
\be
\int \frac{d\lambda}{2\pi} e^{i\lambda x} \langle P S_T |\bar{\psi}_j(0)[0,\lambda n]\psi_i(\lambda n)| P S_T \rangle =  \frac{M_N}{2}(\gamma_5\slashed{S}_T)_{ij} g_T(x) + \dots\,. \label{gtdef}
\ee
while $G_F(x_1,x_2)$ and $\tilde{G}_F(x_1,x_2)$ are the ETQS functions (We follow the notation of Ref.~\cite{Eguchi:2006mc} where explicit definitions can be found). The `kinematical' distributions $\tilde{g}(x)$ and $g_T(x)$ are related through the QCD equation of motion
\be
g_T(x) + \frac{\tilde{g}(x)}{2x} = \int d x' \frac{G_F(x,x') + \tilde{G}_F(x,x')}{x-x'} ~.
\label{eq:gtgtil}
\ee
The hard matrix elements $S_{\mu\nu}^{(0)}(xP)$ and $S^{(1)}_{\mu\nu\alpha}(x_1P,x_2P)$ are computable in perturbation theory. As observed in \cite{Benic:2019zvg}, the first nonzero contribution to $S^{(0)}$ appears at two loops, $S^{(0)}\propto \alpha_s^2$, and $S^{(1)}$ is obtained from $S^{(0)}$ by attaching an extra gluon in all possible ways.  Two representative diagrams contributing to $S^{(0)}$ are shown in Fig.~\ref{fig:quark}. The crosses denote on-shell lines that lead to an imaginary phase via the Cutkosky rules.  We note that Ref.~\cite{Ratcliffe:1985mp} arrived at essentially the same structure as  (\ref{eq:wtot}), but did not specify the hard kernels $S^{(0,1)}$. 

We shall compute $w_{\mu\nu}$ in the WW approximation, namely, by systematically neglecting genuine twist-three distributions everywhere. This is a consistent approximation in the sense that it preserves both QED and QCD gauge invariance. In this approximation, we may write\footnote{ $\Delta \bar{q}(x)$ is formally related to the operator definition of $g_T(x)$ in the negative support region $0>x>-1$. We have checked that the antiquark contribution can be effectively included via the replacement (\ref{eq:gTww}) in the physical region $1>x>0$ using the same hard kernel.}   
\be
g_T(x) \to  \int_x^1 \frac{dx'}{x'} (\Delta q(x')+\Delta\bar{q}(x))\,, \qquad \tilde{g}(x) \approx -2xg_T(x)\,,
\label{eq:gTww}
\ee 
where $\Delta q(x)$ and $\Delta\bar{q}(x)$ are the standard twist-two polarized  quark and antiquark distributions.  Moreover, the first two lines in \eqref{eq:wtot} can be combined into
\be
w_{\mu\nu} \approx \frac{M_N}{2}\int dx g_T(x)S_T^\alpha\left( \frac{\pd}{\pd k_T^\alpha}{\rm Tr}[\gamma_5 \slashed{k} S_{\mu\nu}^{(0)}(k)]\right)_{k = x P}\,.
\label{eq:wsmalww}
\ee 
 We thus arrive at the structure  mentioned in the introduction,
\beq
d\Delta \sigma \sim (\Delta q(x)+\Delta \bar{q}(x)) \otimes H \otimes D_1(z)\,. \label{above}
\eeq
The above formulas hold for each quark flavor.  In practice, we must sum over flavors weighted by the quark electromagnetic charge. In physical cross sections, we thus apply  $g_T(x) \to \sum_f e_f^2 g_{Tf}(x)$ where $g_{Tf}$ is given by (\ref{eq:gTww}) for each quark flavor $f$. 

Let us compare (\ref{above}) with the conventional contribution from the ETQS function \cite{Eguchi:2006qz,Ji:2006br,Eguchi:2006mc}  which schematically reads\footnote{ To avoid confusion, we note that the $G_F$ and $\tilde{G}_F$ pieces in (\ref{eq:wtot}) are not the conventional ETQS contribution quoted here, but rather its ${\cal O}(\alpha_s)$ corrections. }
\beq
d\Delta \sigma_{\rm ETQS} \sim G_{F} (x,x')\otimes H'\otimes D_1(z)\,. \label{etq}
\eeq
Since $H\sim {\cal O}(\alpha_s^2)$ and $H'\sim {\cal O}(\alpha_s)$, naively the former is parametrically  suppressed by a factor of $\alpha_s$. However, the definition $G_F\sim \langle \bar{\psi}gF\psi\rangle$ explicitly contains the coupling $g$ which actually comes from perturbative diagrams. That is, some suppression associated with the coupling $g$ goes into $G_F$ in the convention (\ref{etq}).   As for the soft part, both conceptually and practically, we have a far better grasp of twist-two distributions than twist-three distributions: $\Delta q$ and  $D_1$ have been well constrained thanks to a wealth of experimental data and global QCD analyses, whereas the ETQS functions are still poorly constrained. These considerations  make  (\ref{above}) a new and attractive source of SSA that can be unambiguously calculated without introducing any free parameters. The main goal of this paper is to carry out such a  calculation, both analytically and numerically. But before doing so, let us point out that an entirely analogous contribution exists in the gluon-initiated channel.

\subsection{Gluon-initiated channel}

The gluonic counterpart of $g_T(x)$ for a transversely polarized proton is defined as \cite{Ji:1992eu,Hatta:2012jm,Koike:2019zxc} 
\be
\int \frac{d\lambda}{2\pi} e^{ix\lambda}\langle PS_T |F^{n\alpha}(0)[0,\lambda n] F^{n\beta}(\lambda n)| PS_T\rangle =   i M_N x\calG_{3T}(x) \epsilon^{n \alpha\beta S_T} + \dots\,.
\label{eq:G3T}
\ee
Similar to $g_T(x)$, it can be written as the sum of the WW part and the genuine twist-three part,
\be
{\cal G}_{3T}(x) =\frac{1}{2} \int_x^1 \frac{dx'}{x'} \Delta G(x')+({\rm genuine\ twist\ three})\,, \label{eq:G3Tww}
\ee
where the WW part is related to the polarized (helicity) gluon PDF $\Delta G(x)$, and the genuine twist-three part consists of three-gluon correlators $\sim \langle FFF\rangle$. Their full expressions can be found in \cite{Hatta:2012jm}.

\begin{figure}
  \begin{center}
  \includegraphics[scale = 0.5]{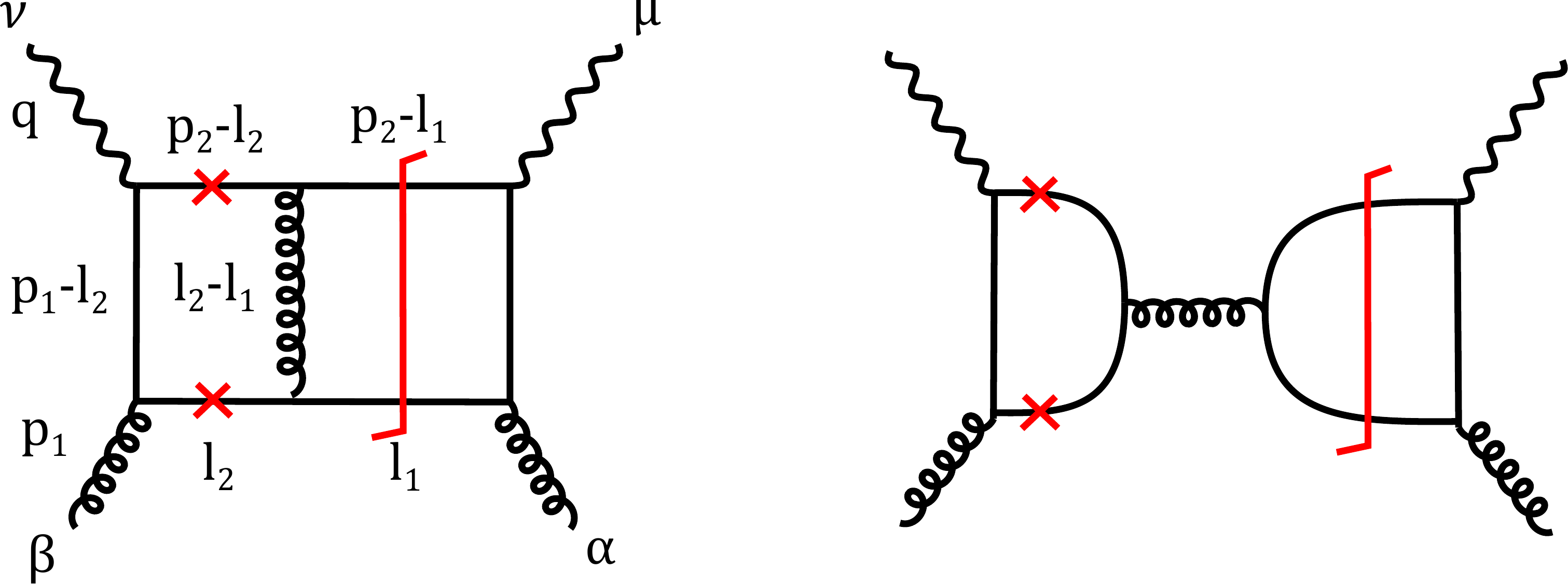}
  \end{center}
  \caption{Prototype two-loop diagrams contributing to SSA in the gluon initiated channel, where the crosses denote the cuts needed for SSA and the vertical line is the final state cut. The right diagram with a $s$-channel gluon does not contribute due to Furry's theorem.}
  \label{fig:gluon}
\end{figure}

The ${\cal G}_{3T}$ distribution appeared in the previous calculation of the double spin asymmetry $A_{LT}$ in proton-proton collisions $p^\rightarrow p^\uparrow \to hX$  \cite{Hatta:2013wsa}. The cross section formula derived in \cite{Hatta:2013wsa} can be straightforwardly adapted to the case of single spin asymmetry in SIDIS $e p^\uparrow \to hX$.
 Writing the hadronic tensor as 
\beq
W_{\mu\nu}^g=\sum_{a=q,\bar{q}}\int \frac{dz}{z^2}D_1^a(z)w_{\mu\nu}^{g,a}\,,
\eeq
we find (see (17) and (25) of \cite{Hatta:2013wsa}) 
\be
\begin{split}
 w^g_{\mu\nu}  =& i M_N \int \frac{dx}{x} \calG_{3T}(x)\epsilon^{n\alpha\beta S_T} S_{\mu\nu}^{(0) \alpha' \beta'} (xP)\omega_{\alpha'\alpha} \omega_{\beta'\beta}\\
& - i M_N \int \frac{dx}{x^2} \tilde{g}(x)\left(g_T^{\beta\lambda} \epsilon^{\alpha P n S_T} - g_T^{\alpha\lambda} \epsilon^{\beta P n S_T}\right)\left( \frac{\pd S_{\mu\nu\alpha \beta}^{(0)} (k)}{\pd k^\lambda}\right)_{k = xP} \\
&-\frac{1}{2}\int \frac{dx_1dx_2}{x_1x_2} M_F^{\alpha\beta\gamma}(x_1,x_2)\frac{S^{(1)\alpha'\beta'\gamma'}(x_1,x_2)}{x_2-x_1}\omega_{\alpha'\alpha}\omega_{\beta'\beta}\omega_{\gamma'\gamma} \,,
\end{split}
\label{eq:wglue}
\ee
where $g_T^{\mu\nu} = g^{\mu \nu} - P^\mu n^\nu - n^\mu P^\nu$ and $\omega^{\mu\nu} = g^{\mu\nu} - P^\mu n^\nu$ are the projectors to the transverse space.  $\tilde{g}(x)$ is again a kinematical function with its precise definition given in \cite{Hatta:2012jm} (see also \cite{Koike:2019zxc} where it is called $\Delta G^{(1)}_T(x)$).  $M_F$ denotes the three-gluon correlators $\langle FFF\rangle$ (see (19) of \cite{Hatta:2013wsa}).  
The hard part also starts at two loops, $S^{(0)}\sim \alpha_s^2$, whose diagrams have the same topology as in the quark-initiated channel. A representative diagram is displayed on the left hand side of Fig.~\ref{fig:gluon}, and $S^{(1)}$ is obtained by attaching a gluon to this diagram in all possible ways. 
The diagram on the right, which is an analog of the right diagram in Fig.~\ref{fig:quark}, does not contribute due to  Furry's theorem.  Note that in the computation of $A_{LT}$ in Ref.~\cite{Hatta:2013wsa}, the imaginary phase comes from the definition of $\Delta G(x)$ for the longitudinally polarized proton. That is,   the non-pole part of the hard kernel was calculated. In the present case, the imaginary phase comes from propagator poles in the hard kernel $S^{\mu\nu}_{\alpha\beta}$, and this is why two-loop diagrams are needed.

 In the WW approximation, we may write 
\be
{\cal G}_{3T}(x) \approx \frac{1}{2} \int_x^1 \frac{dx'}{x'} \Delta G(x')\,, \qquad \tilde{g}(x) \approx x^2 \calG_{3T}(x)\,,
\ee
and neglect the third line of (\ref{eq:wglue}). 
We thus arrive at a new contribution to SSA of the form (\ref{newglue}) which consists only of twist-two distributions.  For light-hadron production, this contribution is suppressed compared to the quark one discussed earlier. However, for SSA in productions of heavy systems such as open charm and $J/\psi$, it is expected to play a more important role. 

\section{Computation of the hard part: Quark-initiated channel} 

In this and the next sections, we embark on an analysis of the two-loop diagrams for the quark and gluon initiated channels, respectively. The calculation is rather involved, especially because nontrivial cancellations of infrared divergences are in demand. In the end, we shall have infrared safe formulas that can be straightforwardly evaluated numerically.

\subsection{Quark-fragmenting channel}

In Fig.~\ref{fig:quark}, either a quark or a gluon in the final state fragments into the observed hadron. For definiteness, we focus on the former process below. The treatment of the latter is basically analogous, and will be included only in the final formulas. 
The hard factor $S^{(0)}_{\mu\nu}$ for the quark-initiated and quark-fragmenting channel explicitly reads
 \cite{Benic:2019zvg}
\be
\begin{split}
S^{(0)\mu \nu}(k) & = -\frac{g^4}{2N_c}(2\pi)\delta\left(\left(k+q-p_q\right)^2\right)\int \frac{d^4 l_2}{(2\pi)^4} (2\pi)\delta\left(l_2^2\right)(2\pi) \delta\left((k + q - l_2)^2\right)\\
&\times \left\{ i A^{\alpha\mu}(k+q-p_q) \bar{M}_{\alpha\beta}(k+q-p_q,l_2)A^{\nu\beta}(l_2)-iA^{\alpha\mu}(l_2) \bar{M}_{\alpha\beta}(l_2,k+q-p_q) A^{\nu\beta}(k+q-p_q)\right\}\,,  
\label{eq:S0}
\end{split}
\ee
where $k$ and $p_q$ are the the momenta of incoming and outgoing quarks, respectively,  and 
\beq
&&\bar{M}_{\alpha\beta}(k+q-p_q,l_2)\nonumber \\ 
&& = \frac{N_c(N_c^2 - 1)}{4}\slashed{p}_q\left[-\frac{V_{\alpha\beta\rho}(k+q-p_q,l_2)\gamma^\rho}{(k+q-p_q - l_2)^2} + \frac{N_c^2 - 1}{N_c^2}\gamma_\alpha \frac{\slashed{k} + \slashed{q}}{(k+ q)^2}\gamma_\beta - \frac{1}{N_c^2}\gamma_\beta \frac{\slashed{p}_q - \slashed{l}_2}{(p_q - l_2)^2}\gamma_\alpha\right](\slashed{k}+\slashed{q} - \slashed{l}_2)\,,
\label{eq:mm}
\eeq
\be
V_{\alpha\beta\rho}(k+q-p_q,l_2) = g_{\alpha\beta}(l_2+k+q-p_q)_\rho + g_{\alpha\rho}(l_2-2(k+q-p_q))_\beta + g_{\rho\beta}(k+q-p_q-2l_2)_\alpha\,,
\ee
\be
A^{\alpha\mu}(k+q-p_q)=
\gamma^\alpha \frac{\slashed{p}_q-\slashed{q}}{(p_q-q)^2}\gamma^\mu +\gamma^\mu\frac{\slashed{k} + \slashed{q}}{(k + q)^2}\gamma^\alpha\,,
\label{eq:Al1}
\ee
\be
A^{\nu\beta}(l_2)=
\gamma^\nu \frac{\slashed{k}-\slashed{l}_2}{(k-l_2)^2}\gamma^\beta +\gamma^\beta\frac{\slashed{k} + \slashed{q}}{(k + q)^2}\gamma^\nu\,.
\label{eq:Al2}
\ee
Equation~(\ref{eq:S0}) represents the sum of $12=2\times 3\times 2$ diagrams, two of which are shown in Fig.~\ref{fig:quark}. One can easily recognize the part of diagrams each piece of Feynman rules corresponds to.  
In \eqref{eq:mm} we have performed a color trace, while the Dirac trace is yet to be done.
There are three $\delta$-functions, one for the unobserved gluon in the final state $\delta((k+q-p_q)^2)$, and the other two come from the poles of internal propagators (denoted by the crosses in Fig.~\ref{fig:quark}).  

Eventually we shall take the collinear limit  $k \to p_1 \equiv xP$ in these expressions and introduce  shorthand notations $p_2 \equiv p_1 + q$ and $l_1 \equiv p_2 - p_q$, the latter being  the momentum of the unobserved gluon in Fig.~\ref{fig:quark}. However, this has to be done with some care because the limit does not commute with the $k_{T}$-derivative acting on the hard kernel  in (\ref{eq:wsmalww}). Let us define 
\be
S^{(0)}_{\mu\nu}(k) = g^4 (2\pi) \delta\left(\left(k+q-p_q\right)^2\right)\int \frac{d^2 \vec{l}_{2T} dl_2^+}{(2\pi)^3 2l_2^+}(2\pi) \delta\left((k + q - l_2)^2\right) \widehat{S}^{(0)}_{\mu\nu}(k)\,,
\label{eq:S0k}
\ee
where $\widehat{S}^{(0)}_{\mu\nu}(k)$ can be read off from \eqref{eq:S0}. 
We first convert the $k_T$-derivatives of the $\delta$-functions to the $x$-derivatives as \cite{Xing:2019ovj}
\beq
&& S_T^\alpha\left(\frac{\pd}{\pd k_T^\alpha} \delta\left(\left(k + q - p_q\right)^2\right)\right)_{k = p_1} = -\frac{p_q \cdot S_T}{p_1 \cdot (p_2 - p_q)} x\frac{\pd}{\pd x}\delta\left((p_2 - p_q)^2\right) = \frac{l_1\cdot S_T}{p_1\cdot l_1}x\frac{\pd}{\pd x}\delta(l_1^2)\,, \label{del1}\\
&& S_T^\alpha\left(\frac{\pd}{\pd k_T^\alpha} \delta\left(\left(k + q - l_2\right)^2\right)\right)_{k = p_1} = -\frac{l_2 \cdot S_T}{p_1 \cdot (p_2 - l_2)} x\frac{\pd}{\pd x}\delta\left((p_2 - l_2)^2\right)\,, \label{del2}
\eeq
and then use integration by parts to shuffle the $x$-derivatives from the $\delta$-functions to the hard factor $\widehat{S}^{(0)}_{\mu\nu}$. From the term $\pd\delta(l_1^2)/\pd x$ in (\ref{del1}), we get a term with $\pd g_T/\pd x$ and a term with
\be
\frac{\pd}{\pd x}\left[x\int \frac{d^2 \vec{l}_{2T} dl_2^+}{(2\pi)^3 2l_2^+}\dots\right]\,.
\label{eq:loopx}
\ee 
From the term  $\pd\delta\left((p_2 - l_2)^2\right)/\pd x$ in (\ref{del2}), we get a term with $\pd\delta(l_1^2)/\pd x$, a term  with $\pd g_T/\pd x$ and a term with 
\be
\frac{\pd}{\pd x}\left\{{\rm Tr}\left[\gamma_5 \slashed{p}_1\widehat{S}^{(0)}_{\mu\nu}(p_1)\right]\right\}\,.
\ee
We further convert $\pd \delta(l_1^2)/\pd x$ to a term with $\pd g_T/\pd x$ and a term 
like  \eqref{eq:loopx}. The two resulting terms with $\pd g_T / \pd x$ cancel. In total, we are led to 
\beq
w_{\mu\nu} &=& \frac{M_N}{2}\int dx g_T(x) S_T^\alpha\left(\frac{\partial}{\partial k_T^\alpha}{\rm Tr}[\gamma_5 \slashed{k} S^{(0)}_{\mu\nu}(k)]\right)_{k = p_1}  \nonumber \\ 
&=& \frac{M_N}{2}\int dx (2\pi)\delta\left(l_1^2\right)
 \Biggl[- x\frac{\pd g_T(x)}{\pd x} \int \frac{d^2\vec{l}_{2T} dl_2^+}{(2\pi)^3 2l_2^+}\frac{l_1\cdot S_T}{p_1\cdot l_1} (2\pi) \delta\left((p_2 - l_2)^2\right) g^4 {\rm Tr}[\gamma_5 \slashed{p}_1 \widehat{S}^{(0)}_{\mu\nu}(p_1)]\nonumber \\ 
&& \qquad \quad -g_T(x)\frac{\pd}{\pd x}\left\{ x \int \frac{d^2\vec{l}_{2T} dl_2^+}{(2\pi)^3 2l_2^+}\left[\frac{l_1\cdot S_T}{p_1\cdot l_1} + \frac{l_2\cdot S_T}{p_1\cdot (p_2 - l_2)}\right] (2\pi) \delta\left((p_2 - l_2)^2\right) g^4 {\rm Tr}\left[\gamma_5 \slashed{p}_1 \widehat{S}^{(0)}_{\mu\nu}(p_1)\right]\right\}\nonumber \\ 
&& \qquad \quad  +x g_T(x)\int \frac{d^2\vec{l}_{2T} dl_2^+}{(2\pi)^3 2l_2^+}\frac{l_2 \cdot S_T}{p_1 \cdot (p_2 - l_2)} (2\pi)\delta\left((p_2 - l_2)^2\right) g^4 \frac{\pd}{\pd x} \left\{{\rm Tr}[\gamma_5 \slashed{p}_1 \widehat{S}^{(0)}_{\mu\nu}(p_1)]\right\}\nonumber \\ 
&& \qquad \quad  + g_T(x) \int \frac{d^2\vec{l}_{2T} dl_2^+}{(2\pi)^3 2l_2^+}(2\pi) \delta\left((p_2 - l_2)^2\right) g^4 S_T^\alpha \left(\frac{\pd}{\pd k_T^\alpha}{\rm Tr}[\gamma_5 \slashed{k} \widehat{S}^{(0)}_{\mu\nu}(k)]\right)_{k = p_1}\Biggr]\,,
\label{eq:lines}
\eeq
for which $l_1$ is fixed through momentum conservation as $l_1 = p_2 - p_q$.

A general proof on both QED and QCD gauge invariance of the hadronic tensor (\ref{eq:wtot}) was given in \cite{Benic:2019zvg}. It was also realized that the first two terms in (\ref{eq:wtot}) contain infrared divergence separately when the momentum $l_2$ becomes collinear to the incoming quark line (see Fig.~\ref{fig:quark}), but the divergences cancel exactly. Now that we have written the original formula in a significantly different form (\ref{eq:lines}), it is a nontrivial task to check  that (\ref{eq:lines}) is gauge invariant and divergence free.  In Appendix A, we show that this is indeed the case, but only after summing all the lines of  (\ref{eq:lines}). Knowing where divergences are hidden in intermediate expressions greatly helps a numerical analysis.

\subsection{Calculation of the hard coefficients}

With \eqref{eq:lw}, \eqref{eq:WDfact} and \eqref{eq:lines}, the polarized cross section \eqref{eq:xspol2} takes the following form
\be
\begin{split}
& \frac{d^6\Delta \sigma}{dx_B dQ^2 d z_f d q_T^2 d\phi d \chi} = \frac{\alpha_{\rm em}^2 \alpha_S^2 M_N}{16\pi^2 x_B^2 S_{ep}^2 Q^2} \sum_k \calA_k \int \frac{d x}{x} \int \frac{dz}{z} (2\pi)\delta\left(\frac{q_T^2}{Q^2} - \left(1-\frac{1}{\hatx}\right)\left(1-\frac{1}{\hatz}\right)\right) \sum_f e_f^2 D_f(z) \\
&\qquad \times \Bigg\{- x^2\frac{\pd g_{T f}(x)}{\pd x} \frac{l_1 \cdot S_T}{p_1 \cdot l_1} \int \frac{d^2 \vec{l}_{2T} d l_2^+}{(2\pi)^3 2 l_2^+}(2\pi) \delta((p_2 - l_2)^2) {\rm Tr}\left[\gamma_5 \slashed{p}_1 \widehat{S}_{\mu\nu}^{(0)}(p_1)\tilde{\calV}^{\mu\nu}_k\right]\\
& \qquad  \qquad -x g_{Tf}(x) \frac{\pd}{\pd x}\Bigg\{x\int \frac{d^2 \vec{l}_{2T} d l_2^+}{(2\pi)^3 2 l_2^+}(2\pi) \delta\left((p_2 - l_2)^2\right)\left[ \frac{l_1 \cdot S_T}{p_1\cdot l_1} +  \frac{l_2 \cdot S_T}{p_1\cdot (p_2 - l_2)}\right] {\rm Tr}\left[\gamma_5 \slashed{p}_1 \widehat{S}_{\mu\nu}^{(0)}(p_1)\tilde{\calV}^{\mu\nu}_k\right]\Bigg\}\\
&\qquad \qquad +xg_{Tf}(x)\int \frac{d^2 \vec{l}_{2T} d l_2^+}{(2\pi)^3 2l_2^+} (2\pi)\delta\left((p_2 - l_2)^2\right) \frac{l_2 \cdot S_T}{p_1\cdot (p_2 - l_2)}  x\frac{\pd}{\pd x}\left\{{\rm Tr}\left[\gamma_5 \slashed{p}_1 \widehat{S}_{\mu\nu}^{(0)}(p_1)\tilde{\calV}^{\mu\nu}_k\right]\right\}\\
&\qquad \qquad +xg_{Tf}(x)\int \frac{d^2 \vec{l}_{2T} d l_2^+}{(2\pi)^3 2 l_2^+}(2\pi) \delta\left((p_2 - l_2)^2\right)\left(\frac{\pd}{\pd k_T^\alpha}{\rm Tr}\left[\gamma_5 \slashed{k} S_T^\alpha \widehat{S}_{\mu\nu}^{(0)}(k)\tilde{\calV}^{\mu\nu}_k\right]\right)_{k = p_1}\Bigg\}\,,
\end{split}
\label{eq:px}
\ee
where the common notations $\hat{x}\equiv x_B/x$ and $\hat{z}\equiv z_f/z$ have been introduced. We have included a flavor summation with explicit charges $e_f^2$, as commented after (\ref{above}). 
Equation~(\ref{eq:px}) contains two $\delta$-function constraints
with the first one $\delta\left(\frac{q_T^2}{Q^2} - \cdots \right)$ originating from $\delta\left(l_1^2\right)$. 
Solving the conditions $l_1^2=(p_2-p_q)^2=0$ and $p_q^2 = (p_2 - l_1)^2 = 0$, we find two roots 
\be
l_{1 (a_1)}^+ = \frac{p_2^+}{2}\left( 1 + a_1 \Delta_1\right) \,, \qquad l_{1 (a_1)}^- = \frac{p_2^-}{2}\left( 1 - a_1 \Delta_1\right)\,, \qquad  \Delta_1 = \sqrt{1 - \frac{4 l_{1 T}^2}{p_2^2}} \,, \qquad a_1 = \pm 1\,.
\label{eq:pqpm}
\ee
Recalling the definition  $z_f = P\cdot P_h/P\cdot q$, we have $p_q^- = p_2^- - l_1^- = \hatz q^-$, whose matching onto \eqref{eq:pqpm} leads to $\Delta_1 = a_1(2\hatz - 1)$. Since $\Delta_1 > 0$, the two roots $l_{1 (a_1)}^+$ effectively split the $z$ integration according to the constraint $a_1 (2 \hatz - 1) > 0$: 
\be
\int dz = \sum_{a_1 = \pm}\int dz \theta\left(a_1 (2\hatz -1)\right)\,.
\label{eq:dzbrok}
\ee
The second $\delta$-function sets $(p_2-l_2)^2=0$, which, together with the condition $l_2^2=0$, give two roots 
\be
l_{2 (a_2)}^+ = \frac{p_2^+}{2}\left( 1 + a_2 \Delta_2\right) \,, \qquad l_{2 (a_2)}^- = \frac{p_2^-}{2}\left( 1 - a_2 \Delta_2\right)\,, \qquad    \Delta_2 = \sqrt{1 - \frac{4 l_{2T}^2}{p_2^2}} \,,\qquad a_2 = \pm 1\,.
\label{eq:l2pm}
\ee 
We are allowed to  perform the $l_2$ integrals as 
\be
\int \frac{d^2 \vec{l}_{2T} d l_2^+}{(2\pi)^3 2 l_2^+}(2\pi) \delta\left((p_2 - l_2)^2\right) = \sum_{a_2 = \pm} \int \frac{d^2 \vec{l}_{2T}}{(2\pi)^2}\frac{1}{2 p_2^2 \Delta_2}\int_{-\infty}^\infty d l_2^+ \delta\left(l_2^+ - l_{2 (a_2)}^+\right) = \frac{1}{32 \pi^2}\sum_{a_2 = \pm} \int_0^1 d\Delta_2 \int_0^{2\pi} d\phi_2\,.
\label{eq:l2int}
\ee
In the last equality, we have switched to the polar coordinate and 
changed the integration variable from $l_{2 T}$ to $\Delta_2$. 
This  facilitates the computation significantly because we do not have to integrate over rational functions involving square roots.

Next, we  compute the Dirac traces using Feyncalc \cite{Shtabovenko:2020gxv} and apply the $x$- and $k_T$-derivatives to the 3rd, 4th and 5th lines of (\ref{eq:px}), which have to be done carefully.  
Note that the $x$-derivative acts outside the $l_2$ integral in the 3rd line. We can only evaluate $\delta\left((p_2 - l_2)^2\right)$ and the evaluation of $\delta\left(l_1^2\right)$ cannot be performed before taking the $x$-derivative.
In the 4th (5th) line the $x$ ($k_T^\alpha$)-derivative is within the $l_2$ integral and so both $p_2 - l_2$ and $l_1$ are put on-shell after the derivatives are taken.

The subsequent integrals over $\phi_2$ and $\Delta_2$ are the most cumbersome part of the entire calculation.   The nontrivial angular dependence comes from the propagator denominators 
\be
\begin{split}
& (l_1 - l_2)^2 = -\left[\frac{p_2^2}{4}(a_1\Delta_1 - a_2\Delta_2)^2 + (\vec{l}_{1T} - \vec{l}_{2T})^2 \right]\,,\\
& (p_2 - l_1 - l_2)^2 = -\left[\frac{p_2^2}{4}(a_1\Delta_1 + a_2 \Delta_2)^2 + (\vec{l}_{1T} + \vec{l}_{2T})^2 \right]\,,
\end{split}
\ee
leading to a $\cos(\phi_1 - \phi_2)$ term, while in the numerator, after taking Dirac traces, we are left with powers of $\cos(\phi_1 - \phi_2)$ as well as linear terms of the forms $\sin(\phi_2 - \Phi_S)$ and $\cos(\phi_2 - \Phi_S)$ arising from $\epsilon^{P n l_2 S_T}$ and $l_2\cdot S_T$, respectively. 
We list the formulas used to carry out such integrals in Appendix \ref{sec:integral}.  
After the $\phi_2$ integration, the $k=1,2,3,4$ terms are proportional to 
\beq
\epsilon^{l_{1T}S_T} = -\frac{1}{z}\epsilon^{P_{hT}S_T} =-q_T \hat{z} \sin(\Phi_S-\chi)\,,
\eeq
where $\epsilon^{12}=-\epsilon^{21}=1$, and the $k=8,9$ terms are proportional to 
\beq
l_1\cdot S_T = -\vec{l}_{1T}\cdot \vec{S}_T = \frac{1}{z}\vec{P}_{hT}\cdot \vec{S}_T = q_T\hat{z}\cos(\Phi_S-\chi)\,.
\eeq

In the individual lines of (\ref{eq:px}), the integral over the modulus $l_{2T}$ has a singularity when $l_{2T} \to 0$ (or when $\Delta_2 \to 1$), and when $a_2 = -1$. However, as we will demonstrate in Appendix A, the total expression is finite because of the QCD Ward identity. Therefore, we first compute the $\phi_2$ integrals for each line separately, sum up the results from all the lines and perform the $l_{2T}$ ($\Delta_2$) integration afterwards.  
One notable feature is that the loop integration yields in principle a different expression for each of the four combinations of the roots $(a_1,a_2)$. However, we have found that after the summation over $a_2$ the results are independent of $a_1$. This is an important consistency check as it effectively ensures that, after all, the split  \eqref{eq:dzbrok}  is not necessary and we are back to the ordinary $z$ integral over a complete domain allowed by kinematics. 

 The above discussion is for the quark-initiated and quark-fragmenting channel. 
We have repeated the whole procedure for the quark-initiated and gluon-fragmenting channel. 
Adding the two pieces, we finally arrive at the total result  
\be
\begin{split}
& \frac{d^6\Delta \sigma}{dx_B dQ^2 d z_f d q_T^2 d\phi d \chi} = \frac{\alpha_{\rm em}^2 \alpha_S^2 M_N}{16\pi^2 x_B^2 S_{ep}^2 Q^2} \sum_k \calA_k \calS_k \int^1_{x_{\rm min}} \frac{d x}{x} \int^1_{z_{\rm min}} \frac{dz}{z}\delta\left(\frac{q_T^2}{Q^2} - \left(1-\frac{1}{\hatx}\right)\left(1-\frac{1}{\hatz}\right)\right) \\
&\times\sum_f e_f^2 \big[D_f(z) x^2\frac{\pd g_{T f}(x)}{\pd x}\dhsig^{qq}_{D k}  + D_f(z) x g_{Tf}(x) \dhsig^{qq}_k + D_g(z) x^2\frac{\pd g_{T f}(x)}{\pd x}\dhsig^{qg}_{D k}  + D_g(z) x g_{Tf}(x) \dhsig^{qg}_k\big]\,,
\end{split}
\label{eq:ddsig}
\ee
where ${\cal S}_k = \sin (\Phi_S-\chi)$ for $k=1,2,3,4$ and ${\cal S}_k=\cos (\Phi_S-\chi)$ for $k=8,9$. Note that we may substitute 
$x\partial g_{Tf}/\partial x \approx  -\Delta q_f(x)$ in the above expression. 
The hard coefficients in the quark-fragmenting ($qq$) channel are given by 
\be
\begin{split}
\dhsig_{D8}^{qq}  &= \frac{\left(N_c^2-1\right) \hatx \hatz}{ 2N_c^2 Q {\left(1-\hatz\right)}^2}\Biggl[ \left(1-\hatz\right) \Bigl(1-\hatx+\hatz-3\hatx \hatz+N_c^2 \left(1-\hatx-\hatz+3\hatx \hatz\right)\Bigr)+2\left(1-2\hatx\right) \hatz \log(\hatz) \Biggr]\,, \\
\dhsig_{D9}^{qq}  &=\frac{\left(N_c^2-1\right) \left(1-\hatx\right) \hatx \hatz  }{2N_c^2 q_T {\left(1-\hatz\right)}^2}  \Biggl[  \left(1-\hatz\right) \left(3N_c^2 \left(1-\hatz\right)+3\hatz-1\right) -2\left(1-2\hatz\right) \log(\hatz)\Biggr]\,,
\end{split}
\ee

\be
\begin{split}
\dhsig_1^{qq}  &=\frac{N_c^2-1  }{2N_c^2 q_T \left(1-\hatz\right) \hatz} \Bigg[(1-\hatz)\bigl(\hatz\hatx  \left(\hatx \left(3+10\hatz\right)-3\left(1+\hatz\right)\right)-1\bigr) \\
& \qquad \qquad +N_c^2 \hatz \left({\hatx}^2 \left(3+2\hatz \left(5\hatz-6\right)\right)-1-3\hatx {\left(1-\hatz\right)}^2\right)+6\hatx \left(2\hatx-1\right) {\hatz}^2 \log(\hatz)\Biggl]\,,\\
\dhsig_2^{qq}  &= \frac{ \left(N_c^2-1\right)\hatx }{{N_c}^2 q_T \left(1-\hatz\right)}  \Biggl[ \left(1-\hatz\right) \bigl(\left(1+N_c^2\right) \left(-1+\hatx\right)+\left(N_c^2-1\right) \left(1-3\hatx\right) \hatz\bigr)+2 \left(2\hatx-1\right) \hatz \log(\hatz)\Biggr]\,, \\
\dhsig_3^{qq} & =\frac{\left(N_c^2-1\right) \hatx}{ 4N_c^2 Q \left(1-\hatx\right) {\left(1-\hatz\right)}^2}     \Biggl[\left(1-\hatz\right) \Bigl\{\left(1-\hatx\right) \left(5\hatx+N_c^2 \left(2-11\hatx\right)\right) \hatz-\left(1+N_c^2\right) (1-\hatx)^2 \\
 & \qquad \qquad -\left(N_c^2-1\right) \left(1+\hatx \left(14\hatx-13\right)\right) {\hatz}^2\Bigr\}-2\hatz \left(1-\hatz-\hatx \left(5-4\hatx-8\left(1-\hatx\right) \hatz\right)\right) \log(\hatz)\Biggr]\,,\\
\dhsig_4^{qq} & =\frac{\left(N_c^2-1\right) \hatx}{  2N_c^2 q_T {\left(1-\hatz\right)}^2}  \Biggl[\left(1-\hatz\right) \Bigl\{3{\left(1-\hatz\right)}^2+\hatx \left(-3+\left(5-4\hatz\right) \hatz\right) \\
&  \qquad \qquad -N_c^2 \left(1-\hatz\right) \left(2-3\hatz+\hatx \left(-2+4\hatz\right)\right)\Bigr\}-2\left(\hatx-{\left(1-\hatz\right)}^2-2\hatx \left(1-\hatz\right) \hatz\right) \log(\hatz)\Biggr]\,,\\
\dhsig_8^{qq}& =\frac{ (N_c^2-1)\hatx }{ 4N_c^2 Q \left(1-\hatx\right) {\left(1-\hatz\right)}^2} \Biggl[ \left(1+N_c^2\right) {\left(1-\hatx\right)}^2+\left(1-\hatx\right) \left(5-10\hatx+N_c^2 \left(6\hatx-5\right)\right) \hatz\\
&  \qquad \qquad  -\left(9+\hatx \left(15\hatx-26\right)-N_c^2 \left(7+9\left(\hatx-2\right) \hatx\right)\right) {\hatz}^2 \\
&  \qquad \qquad  -\left(N_c^2-1\right) \left(3+\hatx \left(4\hatx-9\right)\right) {\hatz}^3-2\hatz \left(\hatx \left(7-4\hatx-2\hatz\right)+\hatz-3\right) \log(\hatz)\Biggr]\,,\\
\dhsig_9^{qq} & =\frac{\left(N_c^2-1\right) \hatx }{ 2N_c^2 q_T {\left(1-\hatz\right)}^2}  \Biggl[ \left(1-\hatz\right) \Bigl(2N_c^2 \left(1-\hatx \left(1-\hatz\right)-2\hatz\right) \left(1-\hatz\right)-3+\hatz \left(9-4\hatz\right)+\hatx \left(3-\left(7-2\hatz\right) \hatz\right)\Bigr) \\
&   \qquad \qquad   -2\left(1+\left(\hatz-3\right) \hatz+\hatx \left(2\hatz-1\right)\right) \log(\hatz)\Biggr]\,,\\
\end{split}
\ee

and in the gluon-fragmenting ($qg$) channel by

\be
\begin{split}
\dhsig_{D8}^{qg} & =\frac{ \left(N_c^2-1\right) \hatx }{2N_c^2 Q \hatz}   \Biggl[\hatz \left(-2+\hatz-N_c^2 \hatz+\hatx \left(4-3\hatz+N_c^2 \left(-2+3\hatz\right)\right)\right)-2\left(1-2\hatx\right) \left(1-\hatz\right) \log(1-\hatz)\Biggr]\,,\\
\dhsig_{D9}^{qg} & =\frac{ \left(N_c^2-1\right) \left(1-\hatx\right) \hatx \left(1-\hatz\right)  }{2{N_c}^2 q_T {\hatz}^2}  \Biggl[ \hatz \left(2+3\left(N_c^2-1\right) \hatz\right)+\left(2-4\hatz\right) \log(1-\hatz)\Biggr]\,,
\end{split}
\ee

\be
\begin{split}
\dhsig_1^{qg} & = \frac{N_c^2-1}{2N_c^2 q_T {\hatz}^2} \Biggl[\hatz \left(1+{\hatx}^2 \left(-13+23\hatz-10{\hatz}^2\right)+3\hatx \left(2-3\hatz+{\hatz}^2\right)\right) \\
& \qquad \qquad -N_c^2 \left(1-\hatz\right) \left(-1-3\hatx {\hatz}^2+{\hatx}^2 \left(1-8\hatz+10{\hatz}^2\right)\right)-6\hatx \left(-1+2\hatx\right) {\left(1-\hatz\right)}^2 \log(1-\hatz)\Biggr]\,,\\
\dhsig_2^{qg} & = \frac{\left(N_c^2-1\right) \hatx \left(1-\hatz\right)  }{N_c^2 q_T {\hatz}^2}  \Biggr[\hatz \Bigl(2-\hatz+N_c^2 \hatz-\hatx \left(4-3\hatz+{N_c}^2 \left(-2+3\hatz\right)\right)\Bigr)+2\left(1-2\hatx\right) \left(1-\hatz\right) \log(1-\hatz)\Biggr]\,,\\
\dhsig_3^{qg} & = -\frac{\left(N_c^2-1\right) \hatx}{4{N_c}^2 Q \left(1-\hatx\right) {\hatz}^2}   \Biggl[\hatz \Bigl\{\left(-1+\hatz\right) \left(2+\left(N_c^2-1\right) \hatz\right)+\hatx \left(10-24\hatz+13{\hatz}^2+N_c^2 \left(-4+16\hatz-13{\hatz}^2\right)\right)    \\
 & \qquad \qquad   +{\hatx}^2 \left(-8+23\hatz-14{\hatz}^2+N_c^2 \left(4-17\hatz+14{\hatz}^2\right)\right)\Bigr\} \\ 
& \qquad \qquad  +2\left(1-\hatz\right) \left(-1+\hatx \left(5-8\hatz\right)+\hatz+{\hatx}^2 \left(-4+8\hatz\right)\right) \log(1-\hatz)\Biggr]\,,\\
\dhsig_4^{qg} & = -\frac{\left(N_c^2-1\right) \hatx \left(1-\hatz\right) }{2{N_c}^2 q_T {\hatz}^3} \Biggl[\hatz \left(2+\left(4N_c^2-3\right) \hatz-3\left(N_c^2-1\right) {\hatz}^2+\hatx \left(-2+\left(3-2{N_c}^2\right) \hatz+4\left(N_c^2-1\right) {\hatz}^2\right)\right) \\ 
&  \qquad \qquad+2\left({\left(1-\hatz\right)}^2-\hatx \left(1-2\hatz+2{\hatz}^2\right)\right) \log(1-\hatz)\Biggr]\,,\\
\dhsig_8^{qg} & = \frac{\left(N_c^2-1\right) \hatx }{4N_c^2 Q \left(1-\hatx\right) {\hatz}^2}  \Biggl[\hatz \Bigl\{6-\left(1+{N_c}^2\right) \hatz+3\left(N_c^2-1\right) {\hatz}^2+\hatx \left(-22+8\hatz+9{\hatz}^2+{N_c}^2 \left(4-9{\hatz}^2\right)\right)   \\ 
& \qquad \qquad  +{\hatx}^2 \left(16-9\hatz-4{\hatz}^2+N_c^2 \left(-4+3\hatz+4{\hatz}^2\right)\right)\Bigr\}+2\left(1-\hatz\right) \left(3+8{\hatx}^2+\hatz-\hatx \left(11+2\hatz\right)\right) \log(1-\hatz)\Biggr]\,,\\
\dhsig_9^{qg} & = \frac{\left(N_c^2-1\right) \hatx \left(1-\hatz\right) }{2N_c^2 q_T {\hatz}^3}     \Biggl[\hatz \Bigl(\left(4-N_c^2\right) \hatz-4-4\left(N_c^2-1\right) {\hatz}^2+\hatx \left(4+3\left(N_c^2-2\right) \hatz+2\left(N_c^2-1\right) {\hatz}^2\right)\Bigr) \\ 
& \qquad \qquad  +2\left(-2+2\hatx+3\hatz-4\hatx \hatz+{\hatz}^2\right) \log(1-\hatz)\Biggr]\,.
\end{split}
\ee

Let us briefly comment on the analytic structure of the above results.  The  
hard kernels depend on the virtual quark propagator 
$1/(p_1-l_1)^2\approx 1/(-2p_1^+l_1^-)$. We have parametrized the fragmenting parton momentum as $p_q=p_2-l_1=P_h/z$
 and $l_1=P_h/z$ in the $q\to q$ and $q\to g$ channels, respectively. Since $P_h^-=z_f q^-$ and $p_2^-=q^-$,  we have 
 $l_1^-=(1-\hat z)q^-$ in the former case, and $l_1^-=\hat z q^-$ in the latter
case. This is why the hard cross sections for the $q\to q$
and $q\to g$ channels contain the factor $1/(1-\hat z)$ and $1/\hat z$, respectively.  
When $z_f\ll 1$, 
both factors $1/(1-\hatz)=z/(z-z_f)$ and $1/\hatz=z/z_f$ become large as $z$ is varied between $z_f$ and 1.
When $z_f\to 1$, only the former becomes large around the endpoint $z\gtrsim z_f$, so 
the $q\to q$ channel  dominates over the $q\to g$ channel. 
This observation will be confirmed in the later numerical analysis.
 The denominator $q_T$ hints that higher-order corrections
will introduce the large Sudakov logarithms $\ln^2(Q/q_T)$ at low $q_T$, whose resummation should
be implemented in principle. This is however  beyond the scope of this work. 

\section{Computation of the hard part: Gluon initiated channel}

The gluon initiated channel is somewhat simpler, since  the right diagram in Fig.~\ref{fig:gluon} does not contribute due to Furry's theorem as already pointed out. We thus consider only four diagrams:  the left diagram in  Fig.~\ref{fig:gluon} and its crossing diagrams with the photon and gluon attachments being interchanged.  
Considering the quark-fragmenting channel for definiteness, we sum the four diagrams and their complex-conjugates  in the form
\be
\begin{split}
 S^{(0)\mu\nu \alpha \beta} (k) & = -\frac{i g^4}{2 (N_c^2 - 1)} (2\pi)\delta\left((k + q - p_q)^2\right)\int \frac{d^4 l_2}{(2\pi)^4} (2\pi) \delta\left(l_2^2\right) (2\pi) \delta\left((p_2 - l_2)^2\right)\\
&\times \left[A^{\alpha\mu}_{lj}(k+q - p_q)\bar{M}_{jikl}(k+q - p_q,l_2) A^{\nu\beta}_{ik}(l_2) - A^{\beta\nu}_{lj}(k+q - p_q)\bar{M}_{jikl}(k+q-p_q,l_2) A^{\mu\alpha}_{ik}(l_2)\right]\,,
\end{split}
\label{eq:Sg2}
\ee
with
\be
\begin{split}
\bar{M}_{jikl}(k+q-p_q,l_2) & = -\frac{N_c^2 - 1}{4 N_c} \frac{1}{(k + q - p_q - l_2)^2}\left[\slashed{p}_q \gamma^\rho (\slashed{p}_2 - \slashed{l}_2)\right]_{ji} \left[\slashed{l}_2\gamma_\rho (\slashed{k} + \slashed{q} - \slashed{p}_q)\right]_{kl}\,, \label{mbar}
\end{split}
\ee
\be
A^{\alpha\mu}(k+q-p_q) = \gamma^\alpha \frac{\slashed{p}_q - \slashed{q}}{(p_q - q)^2}\gamma^\mu + \gamma^\mu \frac{\slashed{p}_q - \slashed{k}}{(p_q - k)^2} \gamma^\alpha\,,
\ee
\be
A^{\nu\beta}(l_2) = \gamma^\nu \frac{\slashed{p}_1 - \slashed{l}_2}{(p_1 - l_2)^2}\gamma^\beta + \gamma^\beta \frac{\slashed{q} - \slashed{l}_2}{(q - l_2)^2} \gamma^\nu \,. \label{twopole}
\ee
Here $k$ represents the initial gluon momentum, $p_q$ is the observed quark, with the unobserved antiquark carrying the momentum $k+q-p_q$ (equal to $l_1 = p_2 - p_q$ in the collinear limit), and $l_2$ is the loop momentum.  
The derivative $\pd/\pd k^\lambda$ in \eqref{eq:wglue} can be performed along the steps analogous to the case of the quark initiated channel. Defining
\be
 S^{(0) \mu\nu \alpha \beta} (k) = g^4 (2\pi)\delta\left((k+q - p_q)^2\right)\int \frac{d^2 \vec{l}_{2T}}{(2\pi)^3}\frac{dl_2^+}{2 l_2^+} (2\pi) \delta\left((k + q - l_2)^2\right) \widehat{S}^{(0)\mu\nu\alpha\beta}(k)\,,
\ee
where $\widehat{S}^{(0)\mu\nu\alpha\beta}(k)$ can be read off from (\ref{eq:Sg2}),  
we find
\be
\begin{split}
& \frac{d^6\Delta \sigma}{dx_B dQ^2 d z_f d q_T^2 d\phi d \chi} = \frac{\alpha_{\rm em}^2 \alpha_S^2 M_N}{16\pi^2 x_B^2 S_{ep}^2 Q^2} \sum_k \calA_k \int \frac{d x}{x} \int \frac{dz}{z} (2\pi)\delta\left(\frac{q_T^2}{Q^2} - \left(1-\frac{1}{\hatx}\right)\left(1-\frac{1}{\hatz}\right)\right) \sum_f e_f^2 D_f(z)\\
& \times 2i\Bigg\{ \calG_{3T}(x) \int \frac{d^2 \vec{l}_{2T} d l_2^+}{(2\pi)^3 2 l_2^+}(2\pi) \delta\left((p_2 - l_2)^2\right) \epsilon^{n\alpha\beta S_T}  \widehat{S}_{\mu\nu}^{(0)\alpha'\beta'} (p_1)\omega_{\alpha'\alpha} \omega_{\beta' \beta} \tilde{\calV}^{\mu\nu}_k\\
& + x^2\frac{\pd \calG_{3T}(x)}{\pd x} \frac{l_{1T}^\beta \epsilon^{\alpha P n S_T} - l_{1T}^\alpha \epsilon^{\beta P n S_T}}{p_1 \cdot l_1} \int \frac{d^2 \vec{l}_{2T} d l_2^+}{(2\pi)^3 2 l_2^+}(2\pi) \delta\left((p_2 - l_2)^2\right) \widehat{S}_{\mu\nu\alpha\beta}^{(0)}(p_1)\tilde{\calV}^{\mu\nu}_k \\
& + x\calG_{3T}(x) \frac{\pd}{\pd x}\Bigg\{x\int \frac{d^2 \vec{l}_{2T} d l_2^+}{(2\pi)^3 2 l_2^+}(2\pi) \delta\left((p_2 - l_2)^2\right)\left[\frac{l_{1T}^\beta \epsilon^{\alpha P n S_T} - l_{1T}^\alpha \epsilon^{\beta P n S_T}}{p_1\cdot l_1} + \frac{l_{2T}^\beta \epsilon^{\alpha P n S_T} - l_{2T}^\alpha \epsilon^{\beta P n S_T}}{p_1\cdot (p_2 - l_2)}\right] \widehat{S}_{\mu\nu\alpha\beta}^{(0)}(p_1)\tilde{\calV}^{\mu\nu}_k\Bigg\}\\
& - x\calG_{3T}(x)\int \frac{d^2 \vec{l}_{2T} d l_2^+}{(2\pi)^3 2l_2^+} (2\pi)\delta\left((p_2 - l_2)^2\right) \frac{l_{2T}^\beta \epsilon^{\alpha P n S_T} - l_{2T}^\alpha \epsilon^{\beta P n S_T}}{p_1\cdot (p_2 - l_2)}  x\frac{\pd}{\pd x} \widehat{S}_{\mu\nu\alpha\beta}^{(0)}(p_1)\tilde{\calV}^{\mu\nu}_k\\
& - x\calG_{3T}(x)\int \frac{d^2 \vec{l}_{2T} d l_2^+}{(2\pi)^3 2 l_2^+}(2\pi) \delta\left((p_2 - l_2)^2\right)\left(g_T^{\beta\lambda} \epsilon^{\alpha P n S_T} - g_T^{\alpha\lambda} \epsilon^{\beta P n S_T}\right)\left(\frac{\pd}{\pd k^\lambda} \widehat{S}_{\mu\nu\alpha\beta}^{(0)}(k)\right)_{k = p_1} \tilde{\calV}^{\mu\nu}_k\Bigg\}\,.
\end{split}
\label{eq:pxg2}
\ee
It will be useful to write the 2nd line as
\be
\epsilon^{n\alpha\beta S_T} \widehat{S}^{(0)\alpha'\beta'}_{\mu\nu} (p_1)\omega_{\alpha'\alpha} \omega_{\beta' \beta} = \epsilon^{nP\beta S_T} \widehat{S}^{(0)}_{\mu\nu n \beta} - \epsilon^{n P \alpha S_T} \widehat{S}^{(0)}_{\mu\nu\alpha n}\,.
\ee
Similar to \eqref{eq:px} for the quark-initiated channel, the individual lines in \eqref{eq:pxg2} contain infrared divergences which must be canceled in the sum over all the lines. We will prove this cancellation in Appendix \ref{sec:divg1}.

The hard coefficients can be obtained in complete analogy to the quark initiated channel. We have the same sets of roots $(a_1,a_2)$ as before (see \eqref{eq:pqpm} and \eqref{eq:l2pm}).
As we will show in Appendix~\ref{sec:divg1}, a divergence comes neither from the $p_1 - l_2$ propagator (corresponding to the choice $a_2 = -1$ for the root) nor from the $q-l_2$ propagator (corresponding to the choice $a_2 = 1$) in (\ref{twopole}). 
Therefore, the $l_{2T}$ loop integral is finite, which can be performed analytically. We have also confirmed that, similarly to the previous case,   we obtain an expression independent of the choice of the roots for $l_{1(a_1)}^+$ after the loop integral and after the sum over the $l_{2(a_2)}^+$ roots. All in all, the final result for the cross section can be written in a compact way as
\be
\begin{split}
& \frac{d^6\Delta \sigma}{dx_B dQ^2 d z_f d q_T^2 d\phi d \chi} = \frac{\alpha_{\rm em}^2 \alpha_S^2 M_N}{16\pi^2 x_B^2 S_{ep}^2 Q^2} \sum_k \calA_k \calS_k \int^1_{x_{\rm min}} \frac{d x}{x} \int^1_{z_{\rm min}} \frac{dz}{z}\delta\left(\frac{q_T^2}{Q^2} - \left(1-\frac{1}{\hatx}\right)\left(1-\frac{1}{\hatz}\right)\right) \\
&\qquad \qquad \times\sum_f e_f^2D_f(z) \big[ x^2\frac{\pd \mathcal{G}_{3T}(x)}{\pd x}\dhsig^{gq}_{D k}  +  x \mathcal{G}_{3T}(x) \dhsig^{gq}_k\big]\,,
\end{split}
\label{eq:ddsigg}
\ee
with $x\partial {\cal G}_{3T}/\partial x \approx -\Delta G(x)/2$ in the present approximation. 
The hard coefficients are given explicitly by
\be
\begin{split}
\dhsig_{D8}^{gq} & = \frac{2\left(1-\hatx\right) \hatx}{N_c Q (1-\hatz)^2 \hatz}  \Biggl[(1-\hatx) \Bigl(\hat{z}\log(\hat{z}) -(1-\hatz)\log(1-\hatz)\Bigr) + \hatx\hatz(1-\hatz) (1-2\hatz)  \Biggr]\,,\\
\dhsig_{D9}^{gq} & = \frac{2{\left(1-\hatx\right)}^2 \hatx}{N_c q_T {\left(1-\hatz\right)}^2 {\hatz}^2}    \Biggl[{\left(1-\hatz\right)}^2 \log(1-\hatz) + \hatz^2\log(\hatz) +\hatz(1-\hatz)\bigr(\hatz^2+(1-\hatz)^2\bigr) \Biggr]\,,
\end{split}
\ee

\be
\begin{split}
\dhsig_1^{gq} & = \frac{\left(1-\hatx\right) }{N_c q_T \left(1-\hatz\right) {\hatz}^2} \Biggl[\left(1-2\hatz\right) \left(1+2{\hatx}^2 {\left(1-2\hatz\right)}^2-\left(1-\hatz\right) \hatz-2\hatx \left(1-\left(1-\hatz\right) \hatz\right)\right) \\ &  \qquad \qquad  +6\left(1-\hatx\right) \hatx \bigl(\left(1-\hatz\right) \log(1-\hatz)-\hatz \log(\hatz)\bigr)\Biggr]\,,\\
\dhsig_2^{gq} & =\frac{4\left(1-\hatx\right) \hatx}{N_c q_T \left(1-\hatz\right) {\hatz}^2}   \Biggl[\left(1-\hatz\right) \bigl(\hatx \left(2\hatz-1\right) \hatz+\left(1-\hatx\right) \log(1-\hatz)\bigr)-\left(1-\hatx\right) \hatz \log(\hatz)\Biggr]\,,\\
\dhsig_3^{gq} & = \frac{\hatx }{N_c Q {\left(1-\hatz\right)}^2 {\hatz}^2} \Biggl[\left(1-\hatz\right) \hatz \Bigl(1-2\left(1-\hatz\right) \hatz+\hatx \left(-6+\left(13-12\hatz\right) \hatz\right)+{\hatx}^2 \left(5-12\left(1-\hatz\right) \hatz\right)\Bigr) \\ 
&\qquad \qquad   +\left(1-\hatx\right) \Bigl(\left(1-\hatz\right) \left(1-\hatz+\hatx \left(-3+4\hatz\right)\right) \log(1-\hatz)-\hatz \left(1-\hatz-\hatx \left(1-4\hatz\right)\right) \log(\hatz)\Bigr)\Biggr]\,,\\
\dhsig_4^{gq} & = \frac{2\left(1-\hatx\right) \hatx }{N_c q_T {\left(1-\hatz\right)}^2 {\hatz}^3}  \Biggl[\left(1-\hatz\right) \Bigl\{\hatz \Bigl(\hatx-2\hatx \hatz \left(2+\hatz \left(-3+2\hatz\right)\right)-\left(1-\hatz\right) \left(1+\hatz \left(-3+4\hatz\right)\right)\Bigr)  \\ 
&  \qquad \qquad  -\left(1-\hatx \left(1-\hatz\right)\right) \left(1-\hatz\right) \log(1-\hatz)\Bigr\}-\hatx {\hatz}^3 \log(\hatz)\Biggr]\,,\\
\dhsig_8^{gq} & = \frac{\hatx }{N_c Q {\left(1-\hatz\right)}^2 {\hatz}^2}   \Biggl[\left(-1+\hatz\right) \hatz \left(-3+10\hatx-7{\hatx}^2+\left(-2+\hatx\right) \hatz+2\left(1+2\left(-1+\hatx\right) \hatx\right) {\hatz}^2\right) \\ & \qquad  \qquad +\left(1-\hatx\right) \Bigl(\left(1-\hatz\right) \left(3-3\hatz+\hatx \left(-5+2\hatz\right)\right) \log(1-\hatz)-\hatz \left(1-3\hatz+\hatx \left(-1+2\hatz\right)\right) \log(\hatz)\Bigr)\Biggr]\,,\\
\dhsig_9^{gq} & = \frac{-2\left(1-\hatx\right) \hatx }{N_c q_T {\left(1-\hatz\right)}^2 {\hatz}^3}  \Biggl[\left(2+\hatx \left(-2+\hatz\right)-2\hatz\right) {\left(1-\hatz\right)}^2 \log(1-\hatz)+\hatz^2 \left(1-\hatx-\left(2-\hatx\right) \hatz\right) \log(\hatz) \\ 
& \qquad \qquad \qquad \qquad -\hatz \left(1-\hatz\right) \left(-2+3\hatz+\left(-1+\hatz\right) \left(-2\hatx+\hatx \hatz+2\left(-1+\hatx\right) {\hatz}^2\right)\right)\Biggr]\,.
\end{split}
\ee
In this computation we have chosen the quark to be observed in the final state ($p_q \to P_h/z$) while the antiquark goes unobserved. As a nontrivial check we have verified that taking the antiquark as the observed final state ($l_1 \to P_h/z$) and the quark as the unobserved one, we recover exactly the same hard coefficients. 

\section{Numerical results}

With all the analytical results presented in the previous sections, we are now ready to make predictions for EIC measurements.   
Specifically, we will numerically compute the asymmetries from the following definition
\be
A_{UT}^{\sin(\alpha\phi_h + \beta \phi_S)} = \frac{2 \int_0^{2\pi} d\phi_h d\phi_S \sin(\alpha \phi_h + \beta \phi_S) \left[d\sigma(\phi_h,\phi_S)-d\sigma(\phi_h,\phi_S + \pi)\right]}{\int_0^{2\pi} d\phi_h d\phi_S\left[d\sigma(\phi_h,\phi_S)+d\sigma(\phi_h,\phi_S + \pi)\right]}\,,
\label{eq:moments}
\ee
where $d\sigma(\phi_h,\phi_S)$ is a shorthand for
\be
d\sigma(\phi_h,\phi_S) \equiv \frac{d^6 \sigma}{dx_B dQ^2 d z_f d q_T^2 d\phi d \chi}\,.
\ee
The numerator of \eqref{eq:moments} is proportional to the ${\cal O}(\alpha_s^2)$ polarized cross section we calculated. In terms of the Fourier coefficients (\ref{fourier}), we have 
\be
\begin{split}
\calF_1 & = \alpha_S M_N \calF_0\int \frac{dx}{x}\int \frac{dz}{z}\delta\left(\frac{q_T^2}{Q^2} - \left(1-\frac{1}{\hatx}\right)\left(1-\frac{1}{\hatz}\right)\right) \sum_f e_f^2 \\
&\times \big[(1+\cosh^2\psi)
\left(D_f(z)x g_{Tf}(x)\dhsig_1^{qq} + D_f(z)x \mathcal{G}_{3T}(x)\dhsig_1^{gq} + D_g(z)x g_{Tf}(x)\dhsig_1^{qg}\right)\\
& - 2\left(D_f(z)xg_{Tf}(x)\dhsig_2^{qq} + D_f(z)x\mathcal{G}_{3T}(x)\dhsig_2^{gq} + D_g(z)xg_{Tf}(x)\dhsig_2^{qg} \right)\big]\,,\\
\calF_2 & = \alpha_S M_N \calF_0 \int \frac{dx}{x}\int \frac{dz}{z}\delta\left(\frac{q_T^2}{Q^2} - \left(1-\frac{1}{\hatx}\right)\left(1-\frac{1}{\hatz}\right)\right) \sum_f e_f^2(-\sinh 2\psi)\\
& \times\left[D_f(z) x g_{Tf}(x)\dhsig_3^{qq} + D_f(z) x \mathcal{G}_{3T}(x)\dhsig_3^{gq} + D_g(z)x g_{Tf}(x)\dhsig_3^{qg}\right]\,,\\
\calF_3 & = \alpha_S M_N \calF_0 \int \frac{dx}{x}\int \frac{dz}{z}\delta\left(\frac{q_T^2}{Q^2} - \left(1-\frac{1}{\hatx}\right)\left(1-\frac{1}{\hatz}\right)\right) \sum_f e_f^2 \sinh^2\psi\\ 
&\times\left[\left(D_f(z) x g_{Tf}(x)\dhsig_4^{qq} + D_f(z) x \mathcal{G}_{3T}(x)\dhsig_4^{gq} + D_g(z)x g_{Tf}(x)\dhsig_4^{qg}\right)\right]\,,\\
\calF_4 &= \alpha_S M_N \calF_0 \int \frac{dx}{x}\int \frac{dz}{z}\delta\left(\frac{q_T^2}{Q^2} - \left(1-\frac{1}{\hatx}\right)\left(1-\frac{1}{\hatz}\right)\right) \sum_f e_f^2(-\sinh 2\psi)\\
&\times\Big[D_f(z)x^2 g'_{Tf}(x)\dhsig_{D8}^{qq} + D_f(z)x^2 \mathcal{G}'_{3T}(x)(x)\dhsig_{D8}^{gq} + D_g(z)x^2 g'_{Tf}(x)\dhsig_{D8}^{qg}\\
& + D_f(z)xg_{Tf}(x)\dhsig_8^{qq} + D_f(z)x \mathcal{G}_{3T}(x)\dhsig_8^{gq} + D_g(z)xg_{Tf}(x)\dhsig_8^{qg}\Big]\,,\\
\calF_5 & = \alpha_S M_N \calF_0 \int \frac{dx}{x}\int \frac{dz}{z}\delta\left(\frac{q_T^2}{Q^2} - \left(1-\frac{1}{\hatx}\right)\left(1-\frac{1}{\hatz}\right)\right) \sum_f e_f^2\sinh^2 \psi\\
&\times\Big[D_f(z)x^2 g'_{Tf}(x)\dhsig_{D9}^{qq} + D_f(z)x^2 \mathcal{G}'_{3T}(x)\dhsig_{D9}^{gq} + D_g(z)x^2 g'_{Tf}(x)\dhsig_{D9}^{qg}\\
& + D_f(z)xg_{Tf}(x)\dhsig_9^{qq} + D_f(z)x\mathcal{G}_{3T}(x)\dhsig_9^{gq} + D_g(z)xg_{Tf}(x)\dhsig_9^{qg}\Big]\,,\\
\end{split}
\ee
with the definition 
\beq
{\cal F}_0= \frac{\alpha_{\rm em}^2\alpha_S}{16\pi^2x_B^2S_{ep}^2Q^2}\,,
\eeq
and the abbreviations 
\beq
 \qquad xg'_{Tf}(x)\equiv x\frac{ \partial  g_{Tf}(x)}{\partial x} \approx -\Delta q_f(x)\,, \qquad  x{\cal G}'_{3T}(x)\equiv x \frac{ \partial  {\cal G}_{3T}(x) }{\partial x}\approx -\frac{\Delta G(x)}{2}\,.
\eeq
The integration variables $x$ and $z$  are in the ranges
\beq
1>x>x_{\rm min} \equiv  x_B\left(1+\frac{z_f}{1-z_f}\frac{q_T^2}{Q^2}\right)\,,  \qquad  1>z>z_{\rm min}\equiv z_f\left(1+\frac{x_B}{1-x_B}\frac{q_T^2}{Q^2}\right)\,. \label{range}
\eeq

As for the  unpolarized cross section in the denominator, we use the leading-order ${\cal O}(\alpha_s)$ formula \cite{Meng:1991da}, summarized by Eqs.~(54)-(59) in \cite{Eguchi:2006mc}. The angular decomposition can be cast in the following form
\be
\frac{d^6 \sigma}{dx_B dQ^2 d z_f d q_T^2 d\phi d \chi} = F^{1} + F^{\cos\phi_h}\cos\phi_h + F^{\cos 2\phi_h} \cos 2\phi_h\,,\\
\ee
Since we integrate over the lepton angle (see below), only the first term $F^1$ is relevant with the
explicit expression 
\be
\begin{split}
F^1 & = 
{\cal F}_0 \int \frac{dx}{x}\int \frac{dz}{z}\delta\left(\frac{q_T^2}{Q^2} - \left(1-\frac{1}{\hatx}\right)\left(1-\frac{1}{\hatz}\right)\right)\\
&\times \sum_f e_f^2 \Big\{D_f(z)q_f(x) \left[(1+\cosh^2\psi)\hat{\sigma}_1^{qq} - 2\hat{\sigma}_2^{qq}\right] + D_g(z)q_f(x) \left[(1+\cosh^2\psi)\hat{\sigma}_1^{qg} - 2\hat{\sigma}_2^{qg}\right]\\
& \qquad \qquad + D_f(z)G(x) \left[(1+\cosh^2\psi)\hat{\sigma}_1^{gq} - 2\hat{\sigma}_2^{gq}\right]\Big\}\,,\\
\end{split}
\label{unpol}
\ee
where $G(x)$ is the unpolarized gluon PDF and the summation over $f$ includes both quarks and antiquarks.  
The hard factors are given by
\be
\begin{split}
& \hat{\sigma}_1^{qq} = \frac{N_c^2 - 1}{N_c}\hatx\hatz\left\{\frac{1}{Q^2 q_T^2}\left[\frac{Q^4}{\hatx^2 \hatz^2} + (Q^2 - q_T^2)^2\right] + 6\right\}\,,\\
& \hat{\sigma}_2^{qq} = 2 \hat{\sigma}_4^{qq} = 4\frac{N_c^2 - 1}{N_c} \hatx\hatz\,,
\end{split}
\ee
\be
\begin{split}
& \hat{\sigma}_1^{qg} = \frac{N_c^2 - 1}{N_c}\hatx(1-\hatz)\left\{\frac{1}{Q^2 q_T^2}\left[\frac{Q^4}{\hatx^2 \hatz^2} + \frac{(1-\hatz^2)}{\hatz^2}\left(Q^2 - \frac{\hatz^2}{(1-\hatz)^2}q_T^2\right)^2\right] + 6\right\}\,,\\
& \hat{\sigma}_2^{qg} = 2 \hat{\sigma}_4^{qg} = 4\frac{N_c^2 - 1}{N_c} \hatx(1-\hatz)\,,
\end{split}
\ee
\be
\begin{split}
& \hat{\sigma}_1^{gq} = \hatx(1 - \hatx)\left[\frac{Q^2}{q_T^2}\left(\frac{1}{\hatx^2 \hatz^2} - \frac{2}{\hatx \hatz} + 2\right) + 10 - \frac{2}{\hatx} - \frac{2}{\hatz}\right] \,,\\
& \hat{\sigma}_2^{gq} = 2 \hat{\sigma}_4^{gq} = 8 \hatx(1-\hatx) \,.
\end{split}
\ee
Using  $F^1$ (\ref{unpol}) and the relation (\ref{FF}), we obtain from \eqref{eq:moments} 
\be
A_{UT}^{\sin(\alpha \phi_h + \beta \phi_S)} = \frac{F^{\sin(\alpha\phi_h + \beta\phi_S)}}{F^1}\,.
\label{eq:Auts}
\ee

In practice, we show our results as functions of $P_{hT}$, $z_f$ or $x_B$ by integrating out all the other variables in the numerator and denominator.  
Instead of $Q^2$, it is convenient to use $y = Q^2/(x_B S_{ep})$ so that we have $dQ^2 = x_B S_{ep} dy$ and the relations 
\be
\begin{split}
& 1 + \cosh^2\psi = 2 \frac{1 + (1-y)^2}{y^2}\,,\\
& \sinh^2\psi = 4\frac{1-y}{y^2}\,,\\
& \sinh 2 \psi = 4\frac{(2-y)\sqrt{1-y}}{y^2}\,.
\end{split}
\ee
There are general kinematical constraints on the integration ranges of $x_B$, $z_f$ and $y$.  The condition $x_{\rm min}<1$ (see (\ref{range})) leads to
\be
x_B \leq 1 - \frac{1}{z_f(1-z_f)}\frac{P_{hT}^2}{y S_{ep}}\,.
\label{eq:xBmax}
\ee
 Requiring the upper bound of $x_B$ to be positive, we find a condition on $y$,
\be
y>\frac{1}{z_f(1-z_f)}\frac{P_{hT}^2}{S_{ep}}\,.
\ee 
Similar constraints can be obtained from $z_{\rm min}<1$, which are however not very restrictive because $P_{hT}^2\ll S_{ep}$. In actual experiments, $A_{UT}$ is integrated over conveniently chosen bins in $x_B$, $z_f$ and $y$, and we will follow suit.

Since we are using the leading-order cross sections for both the numerator and denominator, one may ask a legitimate question about the effect of higher order corrections, in particular when $P_{hT}\ll Q$ and the resummation of the Sudakov logarithms is required. While such a procedure is well established for unpolarized cross sections, that for transversely polarized cross sections is poorly understood. On a general ground, we expect that the impact of resummation largely cancels in the ratio, but this has to be checked, and will be left for a future work. As for the scale $\mu$ of the QCD coupling constant $\alpha_S$ (and also of PDFs and FFs), we argue that the lower scale $\mu=P_{hT}$ is more appropriate than the larger one $\mu=Q$ in the typical kinematic region $P_{hT}\ll Q$ we are considering. This is understood simply from the aspect 
of the Sudakov ($k_T$) resummation usually done in the Fourier conjugate impact parameter space $b_T$. The running of the coupling tends to pick up a dominant contribution from the large $b_T$ region under the inverse Fourier transformation (for which some prescription is needed to avoid the Landau pole \cite{Collins:1984kg}). Therefore, the choice of a lower scale $\mu=P_{hT}\sim O(1/b_T)$ fits the above all-order picture better.


The computation is performed with the most recent NNPDF and JAM global fits. For the NNPDF sets we use the helicity PDFs from \cite{Nocera:2014gqa} and FFs from \cite{Khalek:2021gxf}.
For the JAM sets we use the helicity PDFs and FFs from a simulatenous fit in \cite{Ethier:2017zbq}. $g_{Tf}(x)$ and $\calG_{3T}(x)$ are deduced  from the helicity PDFs according to the formulas \eqref{eq:gTww} and \eqref{eq:G3Tww} in the WW approximation. 
The uncertainties in PDFs (FFs) in NNPDF and JAM fits are quantified by the Monte Carlo replica method to generate a variance according to the normal distribution. In all the plots below the band represents a combination of 1-$\sigma$ uncertainty due to the replica method and also uncertainty in the scale choice according to $0.5<\mu/P_{hT}< 2.0$, added in quadrature. Note that the edge $\mu=0.5P_{hT}$ starts at $P_{hT}=2$ GeV.
 

\subsection{Calculation for COMPASS kinematics }

\begin{figure}
  \begin{center}
  \includegraphics[scale = 0.5]{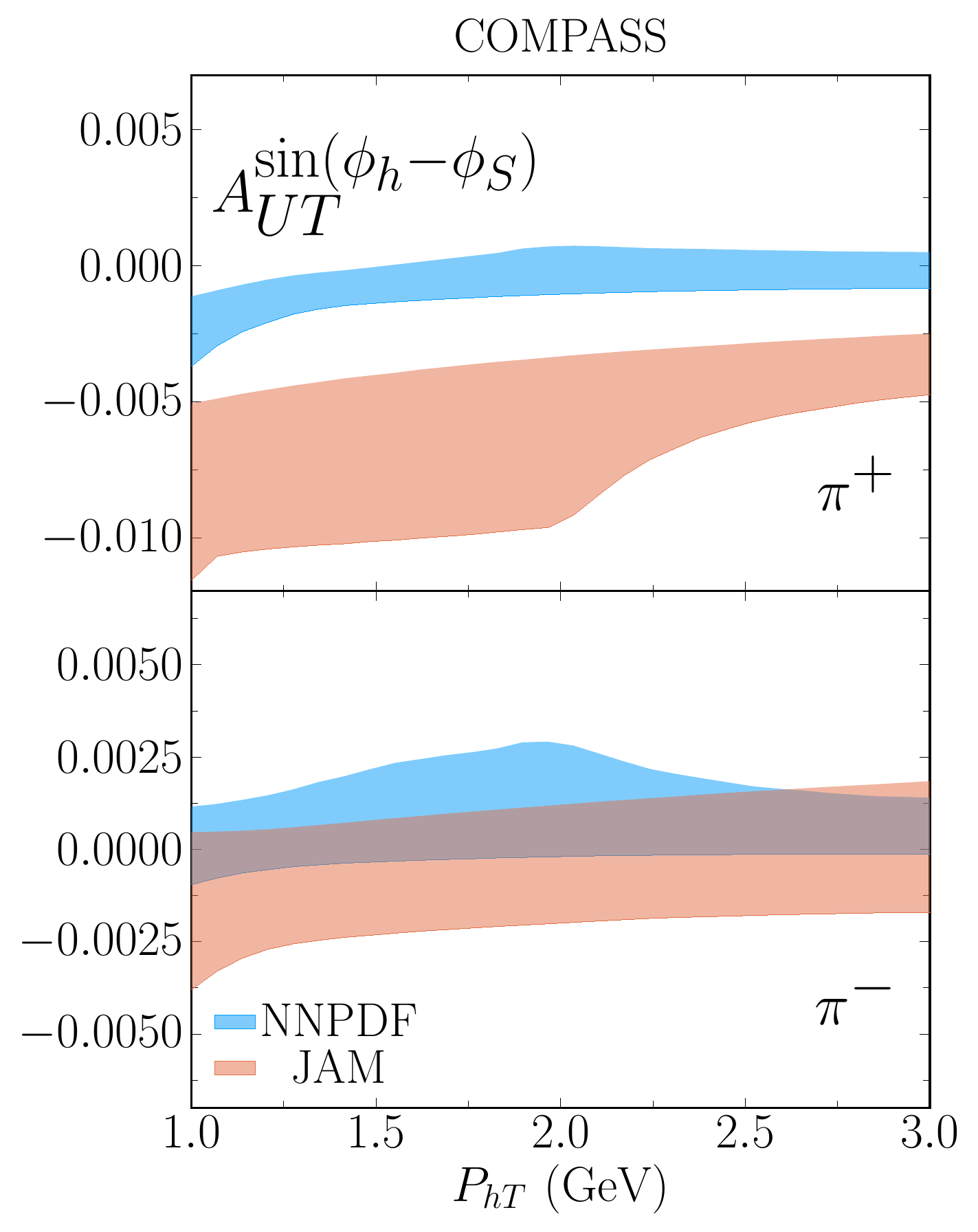}
            \includegraphics[scale = 0.5]{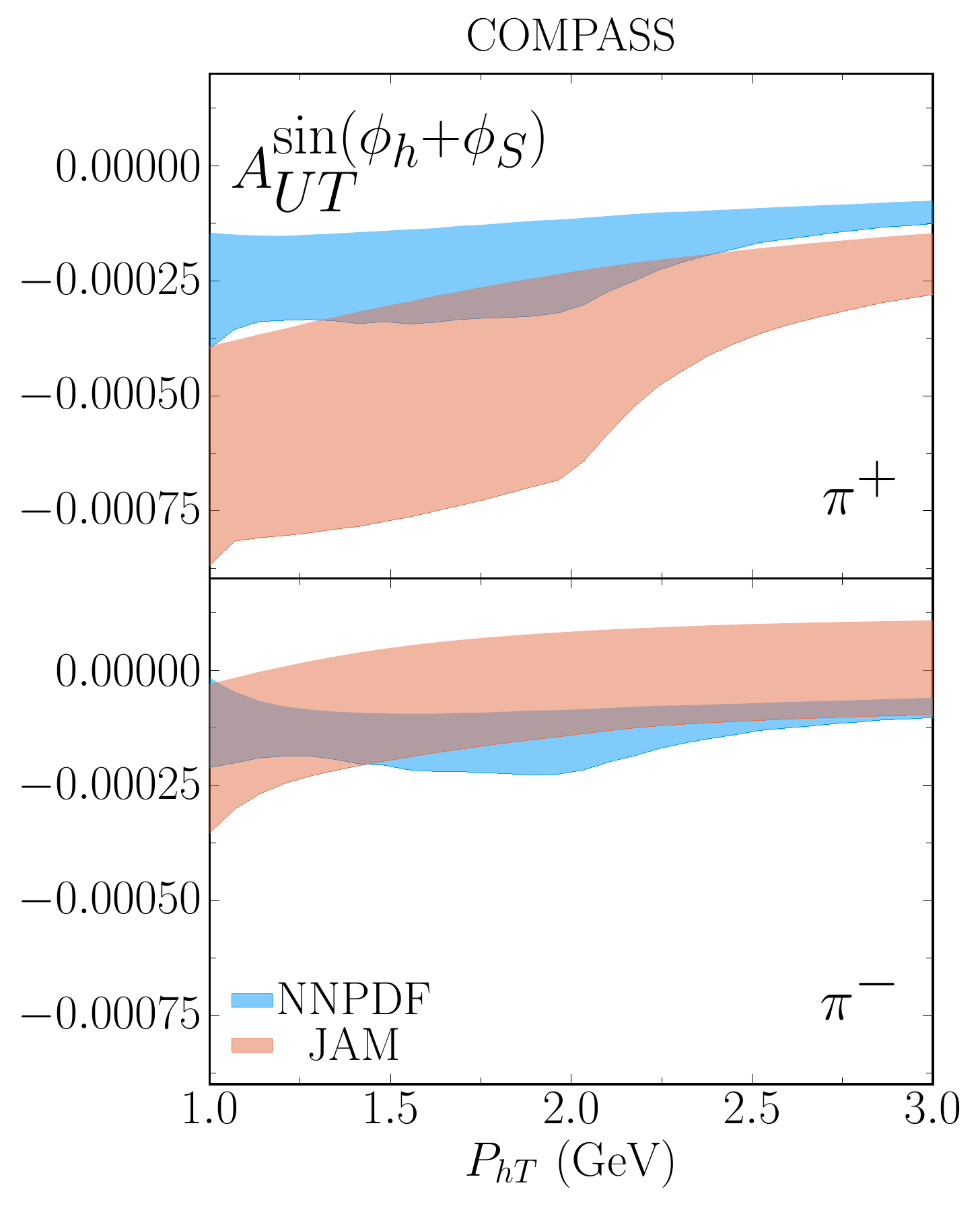}
  \end{center}
  \caption{$P_{hT}$ distributions of Sivers (left) and Collins (right) asymmetries for $\pi^{\pm}$ production at COMPASS.}
  \label{fig:compass}
\end{figure}
Though our approach is most naturally and legitimately applied to the kinematics for EIC, let us first present the results relevant to the COMPASS experiment  \cite{Adolph:2014zba}. Admittedly, the collinear factorization
may not be applicable to the COMPASS kinematics since  most of the data points have $P_{hT}$ below 1 GeV. There is, however,  one published data point with $P_{hT}\approx 1.5$ GeV. We thus only show the $P_{hT}$ distribution for $P_{hT}>1$ GeV, integrating out the other variables over the following coverage   \cite{Adolph:2014zba} 
\be
0.003 \leq x_B \leq 0.7\,, \qquad 0.1 \leq y \leq 0.9 \,, \qquad 0.2 \leq z_f \leq 1\,,
\ee
as well as 
\be
Q^2 > 1 \,\, {\rm GeV}^2,\qquad W^2 > 25 \,\, {\rm GeV}^2\,,
\ee
where $W^2 = (q + P)^2 = Q^2(1 - x_B)/x_B$.
With a 160 GeV muon beam colliding   
on a fixed proton target, the center of mass energy is $\sqrt{S_{\mu p}}\approx 17.4$ GeV. 
 The $P_{hT}$ distributions are shown in Fig.~\ref{fig:compass} for both $\pi^+$ and $\pi^-$ productions. 

We see that the Sivers asymmetry for $\pi^+$ is smaller than $0.5\%$ in magnitude using the NNPDF fits and about $\sim 0.5\%-1\%$ in magnitude using the JAM fits. In either case, the sign is  opposite  to the highest $P_{hT}$ COMPASS data point (see the top-right plot in Fig.~9 of \cite{Adolph:2014zba}). Although 
the significant experimental uncertainty makes a meaningful comparison difficult,  
the result does indicate   
the importance of other sources of SSA, such as the ETQS function. However, $P_{hT}\gtrsim 1$ GeV is the borderline between the collinear and TMD approaches. Therefore, our analysis implies that not only the Sivers function but also the new higher-twist contributions found in \cite{Benic:2019zvg}  need to be 
included in the  global determination of nonperturbative inputs in this regime. As for the Collins asymmetry, our result is negligibly small. The data show nonvanishing central values at $P_{hT}=1.5$ GeV (see the top-right plot in Fig.~6 of \cite{Adolph:2014zba}), but they are consistent with zero after the large error bars are taken into account.

\begin{figure}
  \begin{center}
  \includegraphics[scale = 0.5]{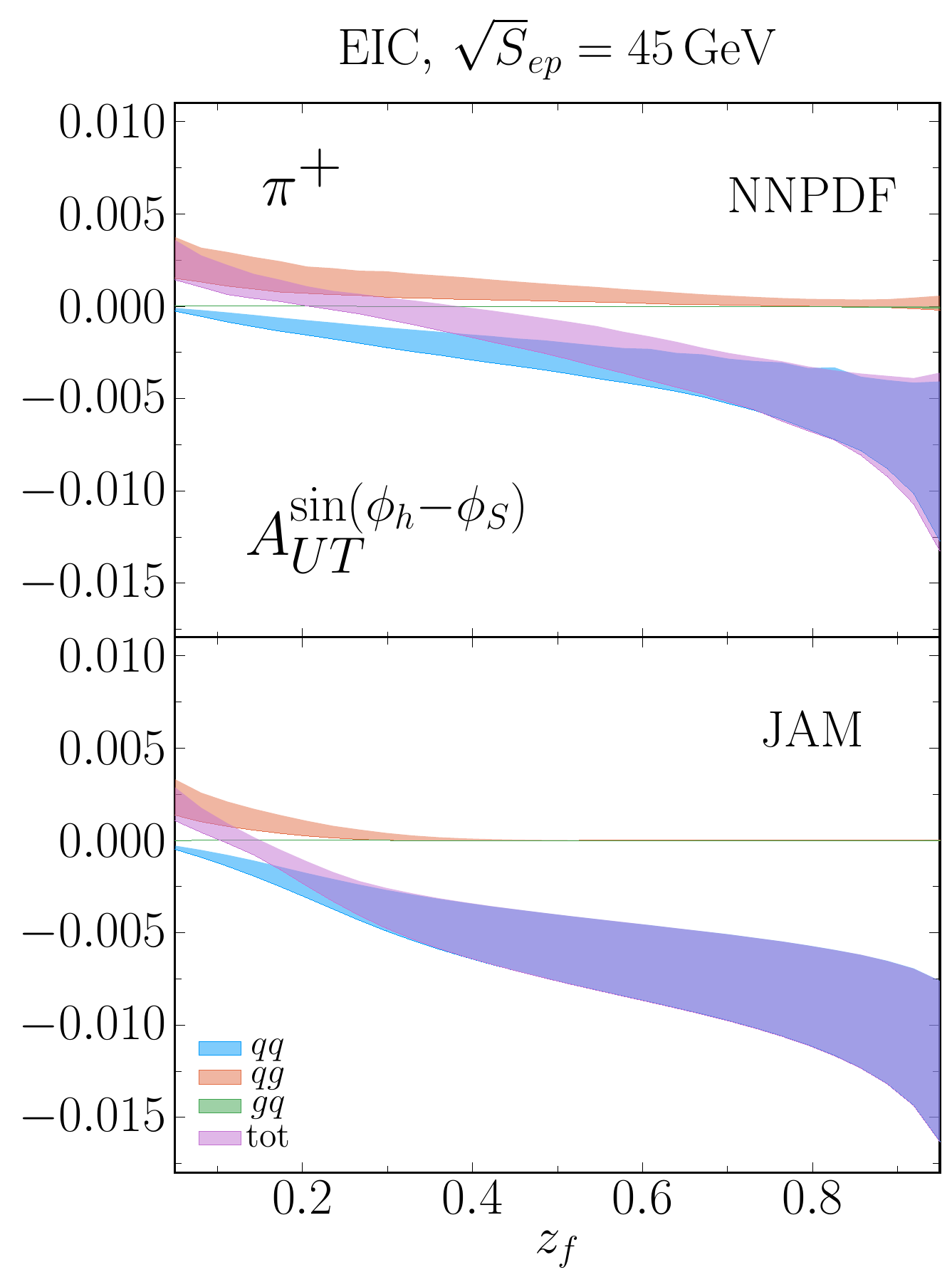}
  \end{center}
  \caption{$z_f$ distributions of the Sivers asymmetry for $\pi^+$ production at $\sqrt{S_{ep}}=45$ GeV differentiated among individual channels.}
  \label{eiczf}
\end{figure}

\begin{figure}
  \begin{center}
  \includegraphics[scale = 0.6]{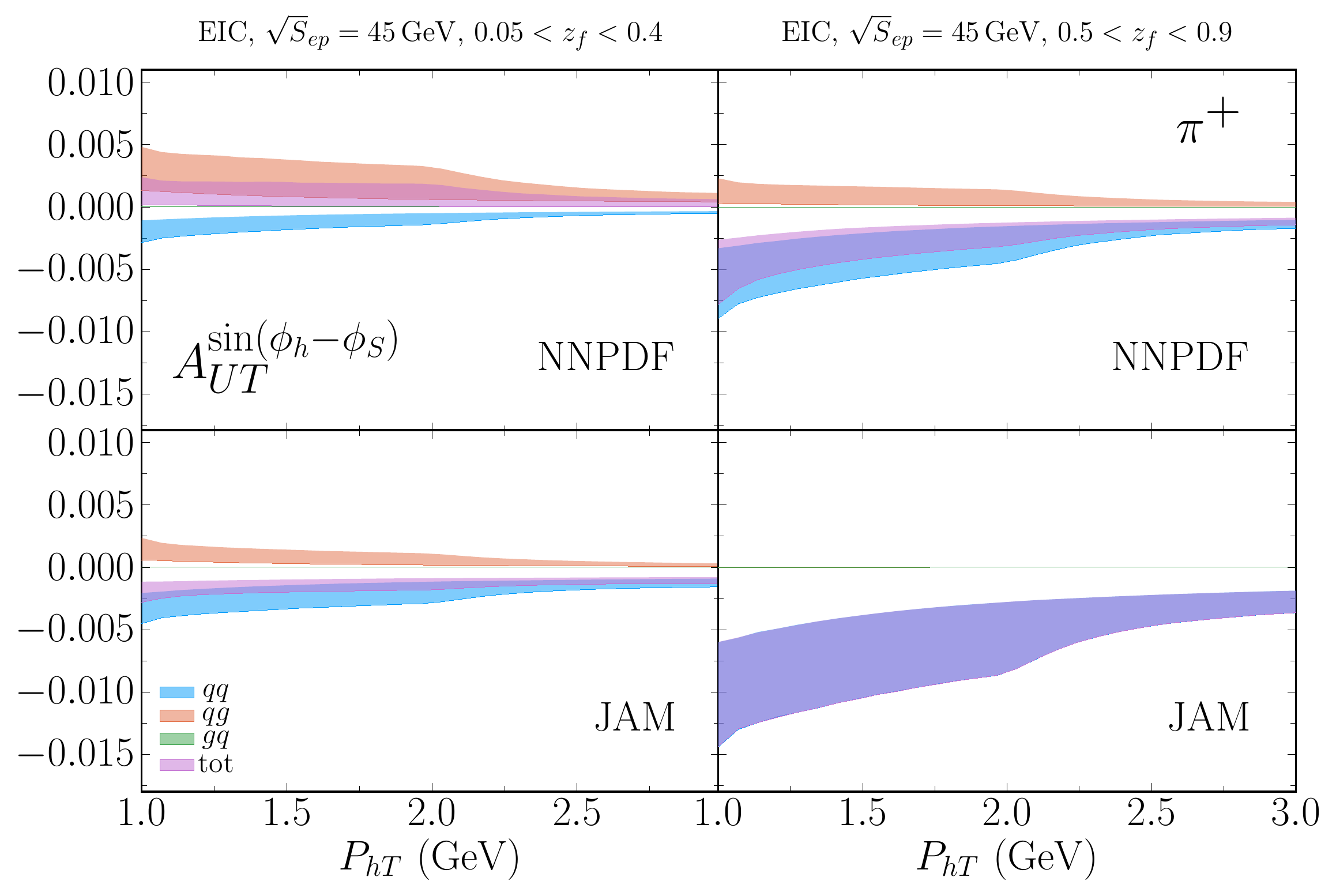}
  \end{center}
  \caption{$P_{hT}$ distributions of the Sivers asymmetry for $\pi^+$ production at $\sqrt{S_{ep}}=45$ GeV in low $z_f$ (left) and high $z_f$ (right) bins.}
  \label{eicpt}
\end{figure}

\subsection{Calculation for EIC kinematics}

We now present our results for the EIC kinematics. 
Figure~\ref{eiczf} shows the $z_f$ distribution of the $\pi^+$ Sivers asymmetry for $\sqrt{S_{ep}}=45$ GeV integrated over the window
$0.1<x_B<0.9$ and $0.01<y<0.95$ and $P_{hT} > 1 \, {\rm GeV}$. The upper bound for the integral over $P_{hT}$ is obtained from \eqref{eq:xBmax} by placing the remaining kinematic variables at their extremal values in the above kinematic window. We also impose the conditions $Q^2>1$ GeV$^2$ and $W^2=(P+q)^2>25$ GeV$^2$. In addition to the total asymmetry (`tot'), respective contributions from different channels ($qq,qg,gq$) are shown.  The asymmetry is largest in the forward region $z_f\to 1$, at most 1.5\% in magnitude, and decreases towards zero as $z_f$ decreases. As we discussed at the end of Section IV, the large $z_f$ region is  dominated by the quark-fragmenting channel, while the gluon-fragmenting channel becomes important at small $z_f$. Since the final state quark and gluon are back-to-back, this explains the sign change for the two channels. A somewhat larger asymmetry is observed from the JAM fit than from the NNPDF fit. This is in fact a general feature seen also for example in Fig.~\ref{fig:compass}, but most directly understood from the $z_f$-distributions in Fig.~\ref{eiczf} where the $qg$ channel contribution is dying off more rapidly as $z_f \to 1$ for the JAM fits. Consequently, the cancellation between the $qq$ and the $qg$ channels is less effective using the JAM fits. The underlying reason is the smaller $g\to \pi^+$ FF in the JAM fit than in the NNPDF fit. 

Figure~\ref{eicpt} gives the $P_{hT}$ distributions of the $\pi^+$ Sivers asymmetry in low $z_f$  ($0.05 < z_f < 0.4$, left) and high $z_f$ ($0.5 < z_f < 0.9$, right) bins. We can again see the role of the $g\to \pi^+$ FF: at low $z_f$ the Sivers asymmetry can even become positive (albeit rather small in magnitude) using the NNPDF fits, while in the large $z_f$ region the $qg$ channel quickly dies off so the JAM fits predict a larger (negative) Sivers asymmetry, around $0.5\%\sim 1.5\%$ in magnitude.

Note that the gluon-initiated ($gq$) channel  is negligibly small, almost invisible in the plots. A closer look reveals that the contribution to $A_N$ from this channel is less than $10^{-3}$ in all the bins we have studied. We have anticipated that the gluon-initiated channel gives a small contribution to light-hadron production. However, the suppression is stronger than expected, and we attempt to explain the reason in the concluding section. 

\begin{figure}
  \begin{center}
  \includegraphics[scale = 0.6]{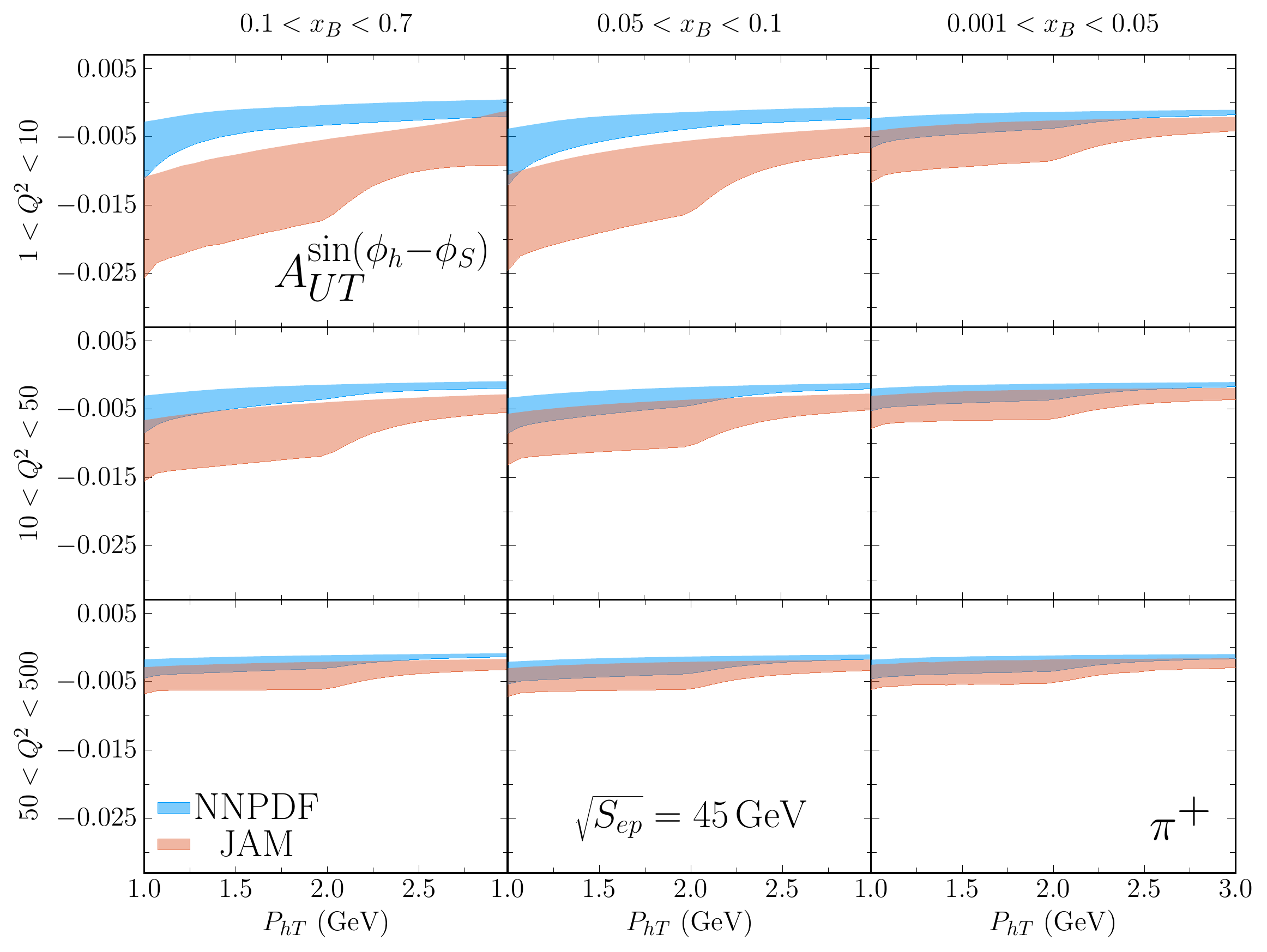}
  \end{center}
  \caption{$P_{hT}$ distributions of the Sivers asymmetry for $\pi^+$ production at $\sqrt{S_{ep}}=45$ GeV for three different $x_B$ and $Q^2$ bins (we have dropped the units in GeV$^2$ on the plot), covering $0.5 < z_f < 0.9$.}
  \label{fig:siverspanelpt}
\end{figure}

\begin{figure}
  \begin{center}
  \includegraphics[scale = 0.6]{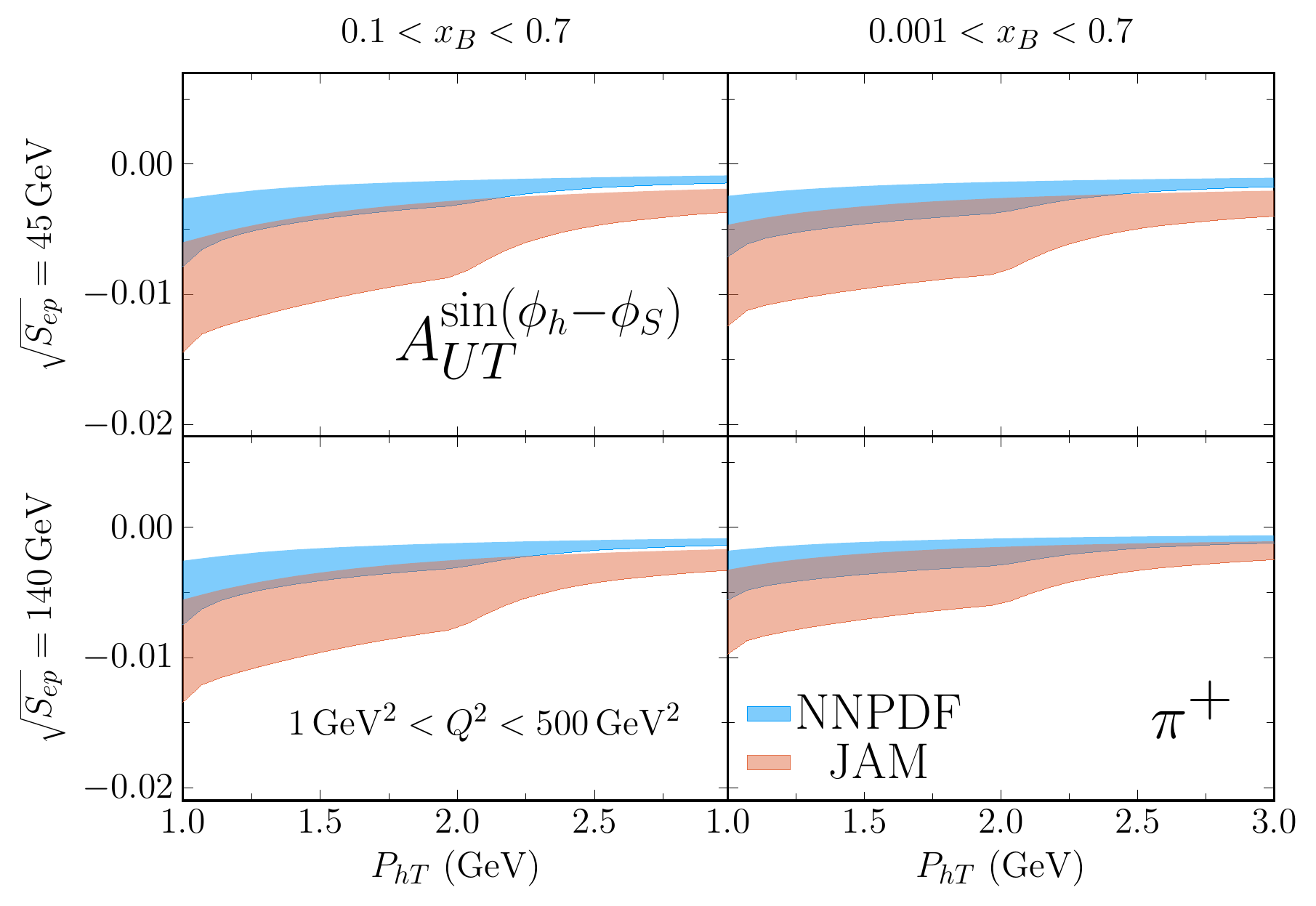}
  \end{center}
  \caption{$P_{hT}$ distributions of the Sivers asymmetry for $\pi^+$ production at $\sqrt{S_{ep}}=45$ GeV and $\sqrt{S_{ep}}=140$ GeV for two different $x_B$ and for $1<Q^2 <500$ GeV$^2$, covering $0.5 < z_f < 0.9$.}
  \label{fig:energypanelpt}
\end{figure}

\begin{figure}
  \begin{center}
  \includegraphics[scale = 0.6]{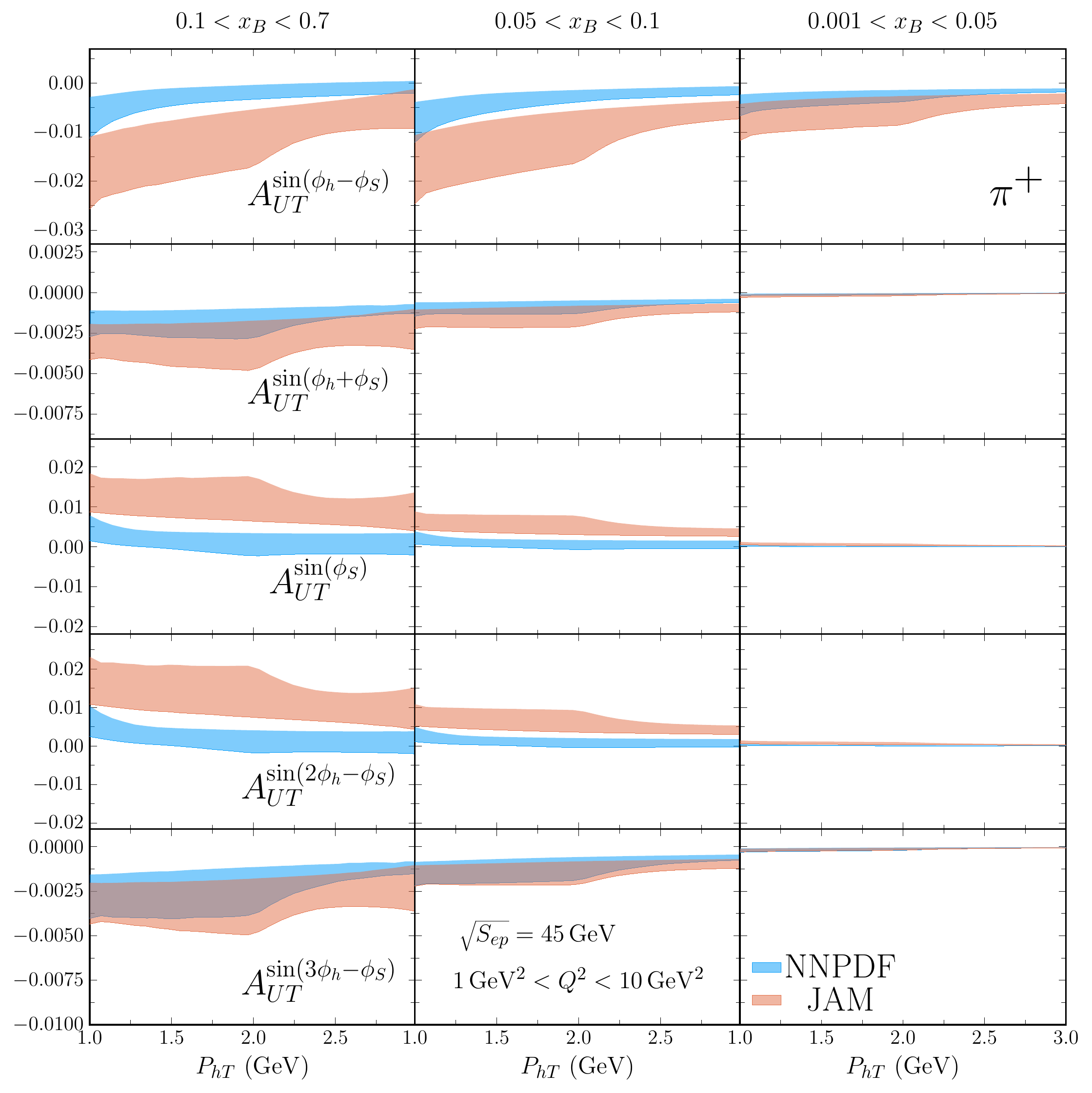}
  \end{center}
  \caption{$P_{hT}$ distributions of all asymmetry moments $A_{UT}$ for $\pi^+$ production at $\sqrt{S_{ep}}=45$ GeV for three different $x_B$ bins integrated over $1<Q^2 <10$ GeV$^2$ and $0.5 < z_f < 0.9$.}
  \label{fig:momentspanelpt}
\end{figure}

Further predictions for the $P_{hT}$ distributions of the $\pi^+$ Sivers asymmetry across three bins in $x_B$ and $Q^2$, using the NNPDF and JAM fits, are exhibited in Fig.~\ref{fig:siverspanelpt}. We find that the Sivers asymmetry can reach up to $2\%$ in magnitude for the JAM fit covering both large $x_B$ ($0.1 < x_B < 0.7$) and moderate $x_B$ ($0.05 < x_B < 0.1$) bins for the lowest $Q^2$ bin. Going from moderate to small $x_B$, the Sivers asymmetry drops to a sub-percent level, as seen in the last $x_B$ bin with $0.001 < x_B < 0.05$. This suppression at small $x_B$ in fact has the same origin as the smallness of the gluon-initiated channel mentioned above (see the discussion in the concluding section).
Figure~\ref{fig:energypanelpt} covers the Sivers asymmetry for two collision energies ($S_{ep}$) and two bins in $x_B$. The results show very mild dropping of the Sivers asymmetry as $\sqrt{S_{ep}}$ is increased from $45$ GeV to the top EIC energy of $140$ GeV (see also  \cite{Matevosyan:2015gwa}). The reason is that the energy dependence mainly comes from the $y$ dependence, which roughly cancels out between the numerator and denominator.   
Compared with an earlier prediction for EIC in the TMD framework at low momentum $P_{hT}<1$ GeV (see Fig.~21 of \cite{Echevarria:2020hpy}), our result for the Sivers asymmetry is similar or somewhat smaller in magnitude but opposite in sign, although a detailed comparison is not possible because there is no overlap in the plotted $P_{hT}$ ranges. This suggests that there are  cancellations   between different mechanisms which may lead to a sign change.  However, we emphasize again that when $P_{hT}\lesssim 1$ GeV, other sources of asymmetries from various twist-three TMDs found  in \cite{Benic:2019zvg}   should be added to the contribution from the Sivers function. 
 
Finally, in Fig.~\ref{fig:momentspanelpt}, we present a full set of moments introduced in \eqref{FF} and \eqref{eq:Auts} for three different bins in $x_B$ and for fixed bins in $Q^2$ and $z_f$ ($1  < Q^2 < 10 \, \mathrm{GeV}^2$, $0.5<z_f<0.9$). 
 We find that two additional moments $A_{UT}^{\sin(\phi_S)}$
and $A_{UT}^{\sin(2\phi_h - \phi_S)}$ reach up to $2\%$ in magnitude in the highest $x_B$ bin $0.7 < x_B < 0.1$. In the TMD framework for low$P_{hT}$, the $\sin(\phi_S)$ and $\sin(2\phi_h-\phi_S)$  asymmetries are known to be generated by various twist-three TMDs \cite{Bacchetta:2006tn}. We have just demonstrated that the $g_T$ distribution  (or its TMD counterpart $g_{1T}$ by extension) is also a potentially significant source of  these asymmetries. Indeed, our prediction 1-2 \% at $P_{hT}=1$ GeV is comparable to  previous  TMD-based calculations \cite{Mao:2014aoa,Wang:2016tix}.

\section{Discussions and Conclusions}

In this paper we have performed the complete analytical and numerical evaluation of the novel two-loop contributions to SSA proportional to the twist-two polarized PDFs $\Delta q(x)$ and $\Delta G(x)$. Our results indicate that, at the EIC, $A_{UT}$ for pions can reach 1-2\%  for the three harmonics $\sin (\phi_h-\phi_S)$, $\sin (\phi_h)$ and $\sin (2\phi_h-\phi_S)$. On the other hand, contributions from the gluon-initiated channel are negligibly small. Since we are dealing with higher-order perturbative  diagrams, we have anticipated that the resulting asymmetry would be small.  However, the  stronger-than-expected suppression we observed, especially in the gluon sector, calls for an explanation.  Parametrically, the asymmetry behaves as 
\beq
A_{UT}=\frac{d\Delta \sigma}{d\sigma} \sim \frac{\alpha_s^2 \frac{M_N}{P_{hT}} (x\Delta q(x)\ {\rm or}\ x\Delta G(x))}{\alpha_s (q(x) \ {\rm or} \ G(x))}\,. \label{our}
\eeq
In addition to the obvious factor of  $\alpha_s$, $A_{UT}$ is suppressed by the smallness of polarized PDFs as compared to unpolarized ones. In particular,  the gluon-initiated channel is expected to be important for $x\ll 1$, but there, $\Delta G(x)\sim xG(x)$ as a rule of thumb.  On top of this, there is a somewhat unexpected extra factor of $x$ in the numerator which comes from the rewriting $\int dx g_T(x)=\int \frac{dx}{x}xg_T(x)$ and $\int dx {\cal G}_{3T}(x)=\int \frac{dx}{x}x{\cal G}_{3T}(x)$. Of course the same factor exists in the unpolarized cross section in the denominator, which is, however, accompanied by $P^+$, and the product  $xP^+=p^+_1$ goes into the hard part and gets absorbed. Therefore, our new contribution, especially in the gluon initiated channel, is strongly suppressed like $\sim x^2$ at low $x$, or in more practical terms, as the selected kinematic bin is sensitive to the low $x_B$ region. This tendency has been clearly shown in Fig.~\ref{fig:momentspanelpt}. 
In the literature, gluon-initiated channels are usually  ignored  in the calculation of $A_{UT}$ for light-hadrons  (see, however, an attempt in $pp$ collisions \cite{Beppu:2013uda}), partly because it is believed to be small, but also because nothing is known about the strength of the three-gluon correlators $\langle FgFF\rangle$. For the first time, we have presented a reliably calculable piece of the gluon initiated contributions, and found very small values.  After all, our main interest in the gluon initiated processes focuses on $A_{UT}$ in heavy systems such as open charm and $J/\psi$.  This will be studied elsewhere.

It is worthwhile to compare (\ref{our}) with the well-known parametric estimate of SSA 

\beq
A_{UT}\sim \frac{\alpha_s m_q}{P_{hT}}\,, \label{kpr}
\eeq
where $m_q\sim $ a few MeV is the current quark mass. This formula has been inferred from the argument in \cite{Kane:1978nd}, and is often quoted in order to emphasize the smallness of SSA in perturbation theory and the necessity to introduce new nonperturbative distributions. The factor of $\alpha_s$ is because one needs loop diagrams such as in Fig.~1 to get an imaginary part, and the factor of $m_q$ is because one needs a helicity flip. However, this suppression by $m_q$ is illusory for the proton initial state. As is clear from the definitions of $g_T$ and ${\cal G}_{3T}$  in (\ref{gtdef}) and  (\ref{eq:G3T}), $m_q$ is replaced by the proton mass $M_N$ (see also a related argument in \cite{Kovchegov:2012ga}). Thus the correct argument in the DIS case would be that, naively $A_{UT}\sim \frac{\alpha_s M_N}{P_{hT}}$ is large, but the coefficient is suppressed  due to   the above-mentioned factor $x^2$, resulting in SSA of about 1\% as we have shown. In SIDIS at $P_{hT}>1$ GeV, this should be  comparable to other nonperturbative origins of SSA.

Precisely measuring $A_{UT}$ in the sub-percent region is challenging at the EIC. Conversely, if the future data on $A_{UT}$ turn out to be consistently  larger than 1\%, most likely genuine twist-three effects are at work. 
But our result must be subtracted when one tries to extract various twist-three distributions. The distinct kinematical features of our contribution, such as the suppression in low $z_f$ and low-$x_B$ regions,  may be useful to isolate this purely perturbative `background'.  
At lower $P_{hT}<1$ GeV, predictions based on the Sivers function are available 
\cite{Matevosyan:2015gwa,Echevarria:2020hpy}. However,  in the TMD regime $P_{hT}<1$ GeV, there are many other sources of the $\sin (\phi_h-\phi_S)$ asymmetry which are unrelated to the Sivers function \cite{Benic:2019zvg}, that must be taken into consideration in order to reliably extract the Sivers function.

\acknowledgements

We thank Shinsuke Yoshida for useful discussions. S.~B. would like to thank for the warm hospitality of the Yukawa Institute for Theoretical Physics, Kyoto University where part of this work was performed for which S.~B. was supported by the JSPS postdoctoral fellowship for foreign researchers under Grant No. 17F117323.
S.~B. and A.~K. are supported by the Croatian Science Foundation (HRZZ) no. 5332 (UIP-2019-04).
The work by Y.~H. is supported by the U.S. Department of Energy, Office of Science, Office of Nuclear Physics, under contract number DE- SC0012704, and also by  Laboratory Directed Research and Development (LDRD) funds from Brookhaven Science Associates.
H.~n.~L is supported by the Ministry of Science and Technology of R.O.C. under Grant No.
MOST-110-2811-M-001-540-MY3.

\appendix

\section{Analysis of infrared divergences: Quark-initiated channel}

In this Appendix  we check that collinear divergences from the $l_{2T}$ integral in (\ref{eq:lines}) cancel. The first step is to understand the QCD Ward identity associated with the $l_2$ gluon.
Starting from Eq.~\eqref{eq:S0}, we can explicitly show that $\slashed{k}S^{(0)}_{\mu\nu}(k)$ satisfies a QCD Ward identity when $A^{\nu\beta}(l_2)$ is replaced by $l_2^\beta$ provided that the momenta $p_q$, $l_1$, $l_2$ and $p_2 - l_2$ are on-shell. The Ward identity of course holds even when we perform the derivative as in \eqref{eq:lines}. Nevertheless, it is important to check this by an explicit computation, starting, not from \eqref{eq:S0}, but from \eqref{eq:lines}.

In the third line of \eqref{eq:lines}, $l_1=k+q-p_q$ is not on-shell, and in the fourth and fifth lines, $p_2 - l_2=k+q-l_2$ is not on-shell either prior to taking the derivative. A generalization of the Ward identity that covers these cases as well is given by
\be
\begin{split}
l_2^\beta\slashed{k} \bar{M}^\mu_{{\color{white}\alpha} \beta}  & = \slashed{k} A^{\alpha\mu}\slashed{p}_q \Bigg\{\frac{N_c(N_c^2 - 1)}{4} \frac{(k + q - p_q)^2}{(k + q - p_q - l_2)^2} \gamma_\alpha (\slashed{k} + \slashed{q} - \slashed{l}_2)\\
&+\frac{N_c(N_c^2 - 1)}{4}\left[-\frac{l_{2\alpha}}{(k + q - p_q - l_2)^2} - \frac{N_c^2 -1}{N_c^2}\gamma_\alpha \frac{\slashed{k} + \slashed{q}}{(k + q)^2}\right](k+ q - l_2)^2\Bigg\}\,,
\label{eq:wardk1}
\end{split}
\ee
where we have introduced
\be
\bar{M}^\mu_{{\color{white}\alpha} \beta} \equiv A^{\alpha\mu} \bar{M}_{\alpha\beta}\,.
\label{eq:Mmubeta}
\ee
In the second line, $l_1$ and $p_2 - l_2$ are on-shell, so \eqref{eq:wardk1} becomes 
\be
l_2^\beta\slashed{p}_1 A^{\alpha\mu} \bar{M}^\mu_{{\color{white}\alpha} \beta} = 0\,.
\label{eq:ward2nd}
\ee
In other words, the second line satisfies the Ward identity by itself.

In the third line, $l_1$ is off-shell and, according to \eqref{eq:wardk1}, we have
\be
l_2^\beta\slashed{p}_1 \bar{M}^\mu_{{\color{white}\alpha} \beta}  = \slashed{p}_1 A^{\alpha\mu}\slashed{p}_q \frac{N_c(N_c^2 - 1)}{4}\left[ \frac{(p_2 - p_q)^2 \gamma_\alpha}{(p_2 - p_q - l_2)^2}\right](\slashed{p}_2 - \slashed{l}_2)\,.\label{4}
\ee
We first take the derivative $\pd /\pd x$, and then take $l_1$ on-shell in the next step. We then need only the derivative of $l_1^2$, 
\be
x\frac{\pd}{\pd x}l_1^2 = x\frac{\pd}{\pd x}(p_2 - p_q)^2 = p_1^\alpha \frac{\pd}{\pd p_1^\alpha} (p_2 - p_q)^2 = 2 p_1\cdot (p_2 - p_q)\,.
\label{eq:xder}
\ee
Multiplying (\ref{4}) by the prefactor in the third line of \eqref{eq:lines}, we find a non-zero result
\be
\begin{split}
-\left[ \frac{l_1 \cdot S_T}{p_1\cdot l_1} +  \frac{l_2 \cdot S_T}{p_1\cdot (p_2 - l_2)}\right] x\frac{\pd}{\pd x} \left(l_2^\beta\slashed{p}_1 \bar{M}^\mu_{{\color{white}\alpha} \beta}\right) &= -2\left[l_1 \cdot S_T +  \frac{(l_2 \cdot S_T)(p_1 \cdot l_1)}{p_1\cdot (p_2 - l_2)}\right]\\ 
&\times\slashed{p}_1 A^{\alpha\mu}(\slashed{p}_2 - \slashed{l}_1)   \frac{N_c(N_c^2 - 1)}{4}\left[\frac{\gamma_\alpha}{(l_1 - l_2)^2}\right] (\slashed{p}_2 - \slashed{l}_2)\,.
\end{split}
\label{eq:ward3rd}
\ee

For the fourth line we first perform $x$-derivative and then put $p_2 - l_2$ on shell in the next step. In practice, this means that for the purpose of taking the derivative, $l_2$ is independent of $x$. The only $x$ dependence then comes from $p_1$. In this case we can again write $x\frac{\pd}{\pd x} = p_1^\alpha \frac{\pd}{\pd p_1^\alpha}$, as in \eqref{eq:xder}, and have
\be
\begin{split}
& \frac{l_2\cdot S_T}{p_1 \cdot (p_2 - l_2)}x\frac{\pd}{\pd x} \left(l_2^\beta\slashed{p}_1 \bar{M}^\mu_{{\color{white}\alpha} \beta}\right)\\
& = 2\frac{(l_2\cdot S_T)(p_1 \cdot l_1)}{p_1 \cdot (p_2 - l_2)} \slashed{p}_1 A^{\alpha\mu} (\slashed{p}_2 - \slashed{l}_1)\frac{N_c(N_c^2 - 1)}{4}\left[\frac{\gamma_\alpha}{(l_1 - l_2)^2}\right] (\slashed{p}_2 - \slashed{l}_2)\\
& - 2 (l_2\cdot S_T)\slashed{p}_1 A^{\alpha\mu} (\slashed{p}_2 - \slashed{l}_1)\frac{N_c(N_c^2 - 1)}{4}\left[ \frac{l_{2\alpha}}{(l_1 - l_2)^2} + \frac{N_c^2 -1}{N_c^2}\gamma_\alpha\frac{\slashed{p}_2}{p_2^2}\right]\,.
\end{split}
\label{eq:ward4th}
\ee
For the fifth line we get
\be
\begin{split}
& S_T^\alpha\left[\frac{\pd}{\pd k_T^\alpha}\left(l_2^\beta \slashed{k} \bar{M}^\mu_{{\color{white}\alpha} \beta}\right)\right]_{k = p_1}  = 2(l_1\cdot S_T)\slashed{p}_1 A^{\alpha\mu} (\slashed{p}_2 - \slashed{l}_1) \frac{N_c(N_c^2 - 1)}{4}\left[\frac{\gamma_\alpha }{(l_1 - l_2)^2}\right](\slashed{p}_2 - \slashed{l}_2)\\
& + 2 (l_2\cdot S_T)  \slashed{p}_1 A^{\alpha\mu} (\slashed{p}_2 - \slashed{l}_1)\frac{N_c(N_c^2 - 1)}{4}\left[\frac{l_{2\alpha}}{(l_1 - l_2)^2} + \frac{N_c^2 -1}{N_c^2}\gamma_\alpha\frac{\slashed{p}_2}{p_2^2}\right]\,.
\end{split}
\label{eq:ward5th}
\ee
Summing up \eqref{eq:ward3rd}, \eqref{eq:ward4th} and \eqref{eq:ward5th} we find that the result is zero.
Therefore, even though the individual lines in \eqref{eq:lines} yield non-zero pieces, the QCD Ward identity is satisfied by their sum.\\

Next we analyze infrared divergences in the $l_{2T}$ integral.  
We have shown in \cite{Benic:2019zvg} that there are no divergences arising from the $l_1 - l_2$ and $p_2 - l_1 - l_2$ propagators in \eqref{eq:mm}. Namely, the divergence in $l_1 - l_2$ would arise when $\vec{l}_{2T} \to \vec{l}_{1T}$, but is cancelled in the symmetric piece of $S^{(0)}_{\mu\nu}$. The potential divergence in $p_2 - l_1 - l_2$, arising when $\vec{l}_{2T} \to - \vec{l}_{1T}$, is explicitly cancelled by the numerator of the respective quark propagator. We therefore devote the remainder of this discussion to the possible divergence from the $p_1 - l_2$ propagator.

From $(p_1 - l_2)^2 = -2 p_1^+ l_2^-$, the propagator denominator has a collinear divergence as $\vec{l}_{2T} \to 0$ ($a_2=+1$ in \eqref{eq:l2pm}). In this limit $l_2^- = l_{2T}^2/2 l_2^+ \to 0$, while from $(p_2 - l_2)^2 = 0$ we get $l_2^+ \to p_2^+$.
Consider the 2nd line in \eqref{eq:lines}. According to the Ward identity \eqref{eq:ward2nd}, we write
\be
\slashed{p}_1 \bar{M}^\mu_{{\color{white}\alpha}+} = -\slashed{p}_1 \frac{l^-_2}{l_2^+}\bar{M}^\mu_{{\color{white}\alpha}-} - \slashed{p}_1 \frac{l^i_2}{l_2^+}\bar{M}^\mu_{{\color{white}\alpha}i}\,.
\ee
We then check whether the piece in the second line of \eqref{eq:lines}
\be
\begin{split}
{\rm Tr}\left[\gamma_5 \slashed{p}_1 \bar{M}^\mu_{{\color{white}\alpha}\beta}\gamma^\nu \frac{\slashed{p}_1 - \slashed{l}_2}{(p_1 - l_2)^2} \gamma^\beta\right] & = {\rm Tr}\left[\gamma_5 \slashed{p}_1 \bar{M}^\mu_{{\color{white}\alpha}-}\gamma^\nu \frac{\slashed{p}_1 - \slashed{l}_2}{(p_1 - l_2)^2} \left(-\frac{l_2^-}{l_2^+}\gamma^+ + \gamma^-\right)\right]\\
& + {\rm Tr}\left[\gamma_5 \slashed{p}_1 \bar{M}^\mu_{{\color{white}\alpha}i}\gamma^\nu \frac{\slashed{p}_1 - \slashed{l}_2}{(p_1 - l_2)^2} \left(-\frac{l_2^i}{l_2^+}\gamma^+ + \gamma^i\right)\right]\,,
\end{split}
\ee
is divergent. The denominator of this expression is $(p_1 - l_2)^2 = - 2p_1^+ l_2^- \sim O(l_{2T}^2)$ as $\vec{l}_{2T} \to 0$. If the numerator is $O(l_{2T})$, the result is finite. There are four terms in the numerator, the first term, containing $l_2^- \gamma^+$, is counted as $l_2^- \sim O(l_{2T}^2)$. The second term, containing $\gamma^-$, vanishes because of $\gamma^-\slashed{p}_1 = 0$. The third term, containing $l_2^i \gamma^+$, is $O(l_{2T})$. The fourth term, containing $\gamma^i$, is at least $O(l_{2T})$ because $\slashed{p}_1$ projects out the $O(1)$ piece as
\be
(\slashed{p}_1 - \slashed{l}_2) \slashed{p}_1 =  \left[(p_1^- - l_2^-)\gamma^+ - \slashed{\boldsymbol{l}}_{2T}\right]\slashed{p}_1\,.
\ee
Therefore, the second line is finite.

In the third line of \eqref{eq:lines} the $x$-derivative acts outside of the $l_2$ integral, so $l_1$ is off-shell. In this case the Ward identity \eqref{eq:ward3rd} leaves us with an additional piece
\be
\left(\frac{l_1 \cdot S_T}{p_1\cdot (p_2 - p_q)} + \frac{l_2 \cdot S_T}{p_1\cdot(p_2 - l_2)}\right)\frac{N_c(N_c^2 - 1)}{4}\frac{l_1^2}{(l_1 - l_2)^2} {\rm Tr}\left[\gamma_5\slashed{p}_1 A^{\alpha\mu} (\slashed{p}_1 - \slashed{l}_1)\gamma_\alpha (\slashed{p}_2 - \slashed{l}_2)\gamma^\nu \frac{\slashed{p}_1 - \slashed{l}_2}{(p_1 - l_2)^2} \frac{\gamma^+}{l_2^+}\right]\,.
\label{eq:div}
\ee
The piece proportional to $l_2\cdot S_T$ is at least $O(l_{2T})$, and must be finite, while the $l_1 \cdot S_T$ piece looks divergent. Applying $x \pd /\pd x$, the divergent remainder from the third line is given by
\be
\frac{N_c(N_c^2 - 1)}{4}\frac{2 l_1 \cdot S_T}{(l_1 - l_2)^2} {\rm Tr}\left[\gamma_5\slashed{p}_1 A^{\alpha\mu} (\slashed{p}_2 - \slashed{l}_1)\gamma_\alpha (\slashed{p}_2 - \slashed{l}_2)\gamma^\nu \frac{\slashed{p}_1 - \slashed{l}_2}{(p_1 - l_2)^2} \frac{\gamma^+}{l_2^+}\right]\,.
\label{eq:div1}
\ee

The analysis of the fourth line can be divided into two parts. In the first part, the derivative $x \pd/\pd x$ hits the $\slashed{p}_1 \bar{M}^\mu_{{\color{white}\alpha}\beta}$ structure, and yields an additional piece according to \eqref{eq:ward4th}, which is finite due to the prefactor $l_2 \cdot S_T$. The second part concerns the piece when $ x \frac{\pd x}{\pd x} = p_1^\alpha\frac{\pd}{\pd p_1^\alpha}$ hits the $p_1 - l_2$ propagator. In this case we get
\be
-p_1^\alpha \frac{\pd}{\pd p_1^\alpha} \left[\frac{\slashed{p}_1 - \slashed{l}_2}{(p_1 - l_2)^2}\right] = \frac{\slashed{p}_1 - \slashed{l}_2}{(p_1 - l_2)^2}\slashed{p}_1 \frac{\slashed{p}_1 - \slashed{l}_2}{(p_1 - l_2)^2}\,,
\ee
whose the denominator is $O(l_{2T}^4)$. Because of the prefactor $(l_2\cdot S_T)$, the numerator should be is least $O(l_{2T}^2)$. We have
\be
(\slashed{p}_1 - \slashed{l}_2)\slashed{p}_1 (\slashed{p}_1 - \slashed{l}_2) = 2 (p_1 \cdot l_2) \slashed{l}_2 = 2 p_1^+ l_2^- \slashed{l}_2\,,
\ee
which is indeed $O(l_{2T}^2)$. Therefore, the fourth line is finite.

Consider finally the fifth line. Similar to the fourth line, we can split the analysis into two parts: the extra piece from the Ward identity contained in \eqref{eq:ward5th} and the remainder. From the Ward identity, the divergence resides in the piece proportional to $l_1\cdot S_T$,
\be
-\frac{N_c(N_c^2 - 1)}{4}\frac{2 l_1\cdot S_T}{(p_2 - p_q - l_2)^2} {\rm Tr}\left[\gamma_5\slashed{p}_1 A^{\alpha\mu} \slashed{p}_q\gamma_\alpha (\slashed{p}_2 - \slashed{l}_2)\gamma^\nu \frac{\slashed{p}_1 - \slashed{l}_2}{(p_1 - l_2)^2} \frac{\gamma^+}{l_2^+}\right]\,,
\label{eq:div2}
\ee
which is of the same form, but with an opposite sign to that of \eqref{eq:div1}. Hence, there is a cancellation between the third and fifth lines.
The remaining divergences are contained in
\be
-{\rm Tr}\left[\gamma_5 \slashed{S}_T \bar{M}^\mu_{{\color{white}\alpha}\beta}\gamma^\nu \frac{\slashed{p}_1 - \slashed{l}_2}{(p_1 - l_2)^2} \gamma^\beta\right] + {\rm Tr}\left[\gamma_5 \slashed{p}_1 \bar{M}^\mu_{{\color{white}\alpha}\beta}\gamma^\nu \frac{\slashed{p}_1 - \slashed{l}_2}{(p_1 - l_2)^2} \slashed{S}_T \frac{\slashed{p}_1 - \slashed{l}_2}{(p_1 - l_2)^2}\gamma^\beta\right]\,.
\label{eq:45div}
\ee
With the Ward identity \eqref{eq:ward2nd}, the divergent piece in the first term is easily deduced as
\be
-{\rm Tr}\left[\gamma_5 \slashed{S}_T \bar{M}^\mu_{{\color{white}\alpha}i}\gamma^\nu \frac{(p_1^+ - l_2^+)\gamma^-}{(p_1 - l_2)^2} \gamma^i\right]\,.
\label{eq:diverg}
\ee
For the second term in \eqref{eq:45div}, we use a similar strategy to obtain
\be
\begin{split}
& {\rm Tr}\left[\gamma_5 \slashed{p}_1 \bar{M}^\mu_{{\color{white}\alpha}-}\gamma^\nu \frac{\slashed{p}_1 - \slashed{l}_2}{(p_1 - l_2)^2} \slashed{S}_T \frac{\slashed{p}_1 - \slashed{l}_2}{(p_1 - l_2)^2}\left(\gamma^- - \frac{l_2^-}{l_2^+}\gamma^+\right)\right]\\
& + {\rm Tr}\left[\gamma_5 \slashed{p}_1 \bar{M}^\mu_{{\color{white}\alpha}i}(p_2 - l_2)\gamma^\nu \frac{\slashed{p}_1 - \slashed{l}_2}{(p_1 - l_2)^2} \slashed{S}_T \frac{\slashed{p}_1 - \slashed{l}_2}{(p_1 - l_2)^2}\left(\gamma^i - \frac{l_2^i}{l_2^+}\gamma^+\right)\right]\,.
\end{split}
\label{eq:div45w}
\ee
Employing
\be
\frac{\slashed{p}_1 - \slashed{l}_2}{(p_1 - l_2)^2} \slashed{S}_T \frac{\slashed{p}_1 - \slashed{l}_2}{(p_1 - l_2)^2} = -2(l_2\cdot S_T)\frac{\slashed{p}_1 - \slashed{l}_2}{(p_1 - l_2)^4} - \frac{\slashed{S}_T}{(p_1 - l_2)^2}\,,
\ee
we find that the divergence is contained only in the transverse part (the second line) in \eqref{eq:div45w}. Because of the multiplication by $\slashed{p}_1$ on the left, the divergent piece is written as
\be
\begin{split}
&\left(-2(l_2\cdot S_T)\frac{\slashed{p}_1 - \slashed{l}_2}{(p_1 - l_2)^4} - \frac{\slashed{S}_T}{(p_1 - l_2)^2}\right)\left(\gamma^i - \frac{l_2^i}{l_2^+}\gamma^+\right)\\
& = 2(l_2\cdot S_T)\frac{\slashed{l}_{2T}}{(p_1 - l_2)^4}\gamma^i + 2(l_2\cdot S_T)\frac{l_2^i}{l_2^+}\frac{p_1^+ - l_2^+}{(p_1 - l_2)^4}\gamma^- \gamma^+ - \frac{1}{(p_1 - l_2)^2}\slashed{S}_T \gamma^i\\
& = -l_{2T}^2 \frac{\slashed{S}_T}{(p_1 - l_2)^4}\gamma^i - \ltwop^2 \frac{S_T^i}{l_2^+}\frac{p_1^+ - l_2^+}{(p_1 - l_2)^4}\gamma^- \gamma^+ - \frac{1}{(p_1 - l_2)^2}\slashed{S}_T \gamma^i\\
& = - \frac{p_1^+ - l_2^+}{p_1^+}\frac{1}{(p_1 - l_2)^2}\left(\slashed{S}_T \gamma^i +  S_T^i \gamma^- \gamma^+\right)\,.
\end{split}
\ee
where in the second line we have performed the angular average, and in the third line we have used $l_{2T}^2 = 2 l_2^+ l_2^-$.
Multiplying the above expression by $\slashed{p}_1$ on the right and using
\be
\slashed{S}_T \gamma^i \gamma^- - S_T^i \gamma^- \gamma^+ \gamma^- = -\gamma^- \gamma^i \slashed{S}_T\,,
\ee
we get
\be
\left(-2(l_2\cdot S_T)\frac{\slashed{p}_1 - \slashed{l}_2}{(p_1 - l_2)^4} - \frac{\slashed{S}_T}{(p_1 - l_2)^2}\right)\left(\gamma^i - \frac{l_2^i}{l_2^+}\gamma^+\right)\slashed{p}_1 = \frac{(p_1^+ - l_2^+)\gamma^-}{(p_1 - l_2)^2}\gamma^i \slashed{S}_T\,.
\ee
The divergent piece of the second term is therefore
\be
{\rm Tr}\left[\gamma_5 \slashed{S}_T \bar{M}^\mu_{{\color{white}\alpha}i}\gamma^\nu \frac{(p_1^+ - l_2^+)\gamma^-}{(p_1 - l_2)^2} \gamma^i\right]\,,
\label{eq:diverg2}
\ee
which is of the same form as \eqref{eq:diverg}, but with an opposite sign.

To conclude this analysis, we have shown that the $l_2$ integral in \eqref{eq:lines} is finite, so we can safely proceed to evaluate it numerically.

\section{Analysis of infrared divergences: Gluon-initiated channel}
\label{sec:divg1}

The gluon initiated channel has two potential sources of infrared divergences coming from the $p_1 - l_2$ and  $q-l_2$ quark propagators in (\ref{twopole}), which are discussed below.
The $p_1 - l_2$ propagator in $A_{\nu\beta}$ causes a divergence as $(p_1 - l_2)^2 = -2p_1^+ l_2^-\propto l_{2T}^2 \to 0$ when the sign $a_2=+1$ is chosen in (\ref{eq:l2pm}). We analyze \eqref{eq:pxg2} line-by-line.  With the multiplication by $\slashed{l}_2$ on the right (see the gamma matrix indices in (\ref{mbar})), the divergent piece in the second line becomes
\be
A_{\nu \beta}\slashed{l}_2 \to \gamma_\nu \frac{\slashed{p}_1 - \slashed{l}_2}{(p_1 - l_2)^2}\gamma_\beta \slashed{l}_2 \to 2p_2^+ \delta_{\beta -} \gamma_\nu \frac{(p_1^+ - l_2^+)\gamma^-}{(p_1 - l_2)^2}\,,
\ee
where we have singled out $\beta = -$. Noting $l_2^-\propto l_{2T}^2$, one can easily check that the divergence is absent when $\beta$ is transverse. 

In the third line, $\beta$ is transverse, so there is no divergence. In the fourth line, $\beta$ is also transverse. The derivative $x \frac{\pd}{\pd x}$ acts outside of the $l_{2T}$ integral, so we are free to evaluate this integral and take the derivative afterwards. We then see that there is no divergence  in the fourth line either.
In the fifth line we take the derivative first. When the derivative hits $A_{\alpha \mu} \bar{M}$, there is no divergence as $\beta$ is transverse. When it hits $A_{\nu \beta}$, we get
\be
x\frac{\pd}{\pd x}A_{\nu\beta} \to -\gamma_\nu \frac{\slashed{p}_1 - \slashed{l}_2}{(p_1 - l_2)^2}\slashed{p}_1 \frac{\slashed{p}_1 - \slashed{l}_2}{(p_1 - l_2)^2} \gamma_\beta \,.
\ee
Because $(\slashed{p}_1 - \slashed{l}_2)\slashed{p}_1 (\slashed{p}_1 - \slashed{l}_2) \sim O(l_{2T}^2)$ and the prefactor going as $O(l_{2T})$ in the fifth line, the numerator in total behaves like $O(l_{2T}^3)$, and the $l_{2T}$ integral is finite.

In the sixth line, when $\pd /\pd k^\lambda$ hits $A_{\alpha\mu} \bar{M}$, the result is finite because $\beta$ is transverse. When it hits $A_{\nu\beta}$, we get
\be
\frac{\pd}{\pd k^\lambda} A_{\nu\beta} \to - \gamma_\nu \frac{\slashed{p}_1 - \slashed{l}_2}{(p_1 - l_2)^2}\gamma_\lambda \frac{\slashed{p}_1 - \slashed{l}_2}{(p_1 - l_2)^2} \gamma_\beta  = 2 l_{2\lambda} \gamma_\nu \frac{\slashed{p}_1 - \slashed{l}_2}{(p_1 - l_2)^4} \gamma_\beta + \gamma_\nu \gamma_\lambda \gamma_\beta \frac{1}{(p_1 - l_2)^2}\,.
\ee
Multiplying it by $\slashed{l}_2$ on the right, we get
\be
\begin{split}
\frac{\pd}{\pd k^\lambda} A_{\nu\beta}\slashed{l}_2 &\to \left( 2 l_{2\lambda} \gamma_\nu \frac{\slashed{p}_1 - \slashed{l}_2}{(p_1 - l_2)^4} + \gamma_\nu \gamma_\lambda  \frac{1}{(p_1 - l_2)^2} \right)\left(2 l_{2\beta} - \slashed{l}_2 \gamma_\beta\right)\\
&\to -2 p_2^+ \delta_{\lambda\beta}\frac{1}{p_1^+}\gamma_\nu\frac{(p_1^+ - l_2^+)\gamma^-}{(p_2 - l_1)^2}\,,
\end{split}
\ee
where we have performed the angular average and inserted $l_{2T}^2 = 2 l_2^+ l_2^-$.

The total contribution then goes as
\be
\begin{split}
& -\frac{1}{x} \epsilon^{nP\alpha S_T} \widehat{S}^{(0)}_{\mu\nu\alpha - } \frac{1}{P^+} - \left(g_T^{\beta\lambda} \epsilon^{\alpha P n S_T} - g_T^{\alpha\lambda} \epsilon^{\beta P n S_T}\right)\left(\frac{\pd}{\pd k^\lambda}\widehat{S}_{\mu\nu\alpha\beta}^{(0)}(k)\right)_{k = p_1}\\
&\to -\epsilon^{n P\alpha S_T} 2 p_2^+  \frac{1}{p_1^+}\gamma_\nu\frac{(p_1^+ - l_2^+)\gamma^-}{(p_1 - l_2)^2}  + \left(g_T^{\beta\lambda} \epsilon^{\alpha P n S_T} - g_T^{\alpha\lambda} \epsilon^{\beta P n S_T}\right) 2 p_2^+ \delta_{\lambda\beta} \frac{1}{p_1^+}\gamma_\nu\frac{(p_1^+ - l_2^+)\gamma^-}{(p_1 - l_2)^2} = 0\,.\\
\end{split}
\ee
Therefore, even though the second and the sixth lines are separately divergent when $l_2^+ \to p_2^+$, $l_2^- \to 0$ and $l_{2T} \to 0$, there is no divergence in their sum.

In addition, the gluon initiated channel has a potential divergence when $l_2^+ \to 0$, $l_2^- \to p_2^-$ and $l_{2T} \to 0$ in the $q - l_2$ quark propagator ($a_2=-1$  in \eqref{eq:l2pm}), since $(q - l_2)^2 = - 2 p_1 \cdot (p_2 - l_2) = - 2p_1^+ (p_2^- - l_2^-)$ with $(p_2 - l_2)^2 = 0$.
The divergent term in the second line is identified as
\be
(\slashed{p}_2 - \slashed{l}_2) A_{\nu \beta} \to 2 p_2^+ \delta_{\beta -} \frac{q^+ \gamma^-}{(q - l_2)^2} \gamma_\nu\,.
\ee
Including the overall prefactors, we are led to
\be
-\epsilon^{n P \alpha S_T} \frac{1}{x} \widehat{S}^{(0)}_{\mu\nu\alpha -} \to \frac{2 p_2^+}{p_1^+} \epsilon^{\alpha P n S_T} \frac{q^+ \gamma^-}{(q - l_2)^2} \gamma_\nu\,.
\label{eq:divg2}
\ee

The third line is finite as before. In the fourth line, $\beta$ is transverse in $\widehat{S}^{(0)}$, but we have a prefactor that now goes as $O(1/l_{2T})$. Since the derivative $x \pd/\pd x$ acts outside the $l_{2T}$ integral, we can first evaluate the $\delta$-function. The relevant piece is given by
\be
\frac{\pd}{\pd x}\left[x\frac{l_{2T}^\beta \epsilon^{\alpha P n S_T} - l_{2T}^\alpha \epsilon^{\beta P n S_T}}{p_1\cdot (p_2 - l_2)}(\slashed{p}_2 - \slashed{l}_2) A_{\nu\beta}\right]\,.
\ee
Because of the condition $(p_2 - l_2)^2 = 0$, $l_2$ also depends on $x$, in addition to $p_1$. In this case we have
\be
x\frac{\pd l_2^\mu}{\pd x} = \frac{1}{2}\frac{p_1^+}{p_2^+} \tilde{l}_2^\mu\,, \qquad \tilde{l}_2^\mu = (2 l_2^+,0,\vec{l}_{2T})\,,
\ee
and calculate the derivative as follows,
\be
\begin{split}
& x\frac{\pd}{\pd x}\left[\left(l_{2T}^\beta \epsilon^{\alpha P n S_T} - l_{2T}^\alpha \epsilon^{\beta P n S_T}\right)(\slashed{p}_2 - \slashed{l}_2) A_{\nu\beta}\right]\\
& = \left[x\frac{\pd}{\pd x}\left(l_{2T}^\beta \epsilon^{\alpha P n S_T} - l_{2T}^\alpha \epsilon^{\beta P n S_T}\right)\right](\slashed{p}_2 - \slashed{l}_2) A_{\nu\beta} + \left(l_{2T}^\beta \epsilon^{\alpha P n S_T} - l_{2T}^\alpha \epsilon^{\beta P n S_T}\right) x\frac{\pd}{\pd x}\left[(\slashed{p}_2 - \slashed{l}_2) A_{\nu\beta}\right]\\
& = \left(l_{2T}^\beta \epsilon^{\alpha P n S_T} - l_{2T}^\alpha \epsilon^{\beta P n S_T}\right)\left[\slashed{p}_1 A_{\nu\beta} + \left(-1 + \frac{1}{2}\frac{p_1^+}{p_2^+}\right)(\slashed{p}_2 - \slashed{l}_2)A_{\nu\beta} - \frac{1}{2}\frac{p_1^+}{p_2^+}\slashed{\tilde{l}}_2 A_{\nu\beta} - \frac{1}{2}\frac{p_1^+}{p_2^+}\gamma_\beta\frac{\slashed{\tilde{l}}_2}{(q-l_2)^2}\gamma_\nu\right]\,.
\end{split}\label{10}
\ee
viewing that $x/p_1\cdot (p_2 - l_2)$ is independent of $x$.
In the fifth line, the $x$-derivative acts inside the $l_2$ integral, so the only effect is
\be
x \frac{\pd}{\pd x} (\slashed{p}_2 - \slashed{l}_2) A^{\nu}_{{\color{white} \nu}\beta} = \slashed{p}_1 A^{\nu}_{{\color{white} \nu}\beta}\,.
\ee
which cancels the first term in the square brackets of the last expression in (\ref{10}). Therefore, we focus on the remaining pieces in the square brackets:
\be
\begin{split}
& \frac{l_{2T}^\beta \epsilon^{\alpha P n S_T} - l_{2T}^\alpha \epsilon^{\beta P n S_T}}{p_1 \cdot (p_2 - l_2)}\left[\left(-1 + \frac{1}{2}\frac{p_1^+}{p_2^+}\right)(\slashed{p}_2 - \slashed{l}_2)A_{\nu\beta} - \frac{1}{2}\frac{p_1^+}{p_2^+}\slashed{\tilde{l}}_2 A_{\nu\beta} - \frac{1}{2}\frac{p_1^+}{p_2^+}\gamma_\beta\frac{\slashed{\tilde{l}}_2}{(q-l_2)^2}\gamma_\nu\right]\\
& \to \left[ 2\left(1 - \frac{p_2^+}{p_1^+}\right) \epsilon^{\alpha P n S_T} - \epsilon^{\beta P n S_T} \gamma^\alpha \gamma_\beta \right] \frac{q^+ \gamma^-}{(q - l_2)^2}\gamma_\nu\,.
\end{split}
\ee 

Finally, the relevant piece in the sixth line is given by
\be
\begin{split}
\left(g_T^{\beta\lambda} \epsilon^{\alpha P n S_T} - g_T^{\alpha\lambda} \epsilon^{\beta P n S_T}\right)\left\{\frac{\pd}{\pd k^\lambda} \left[\left(\slashed{k} + \slashed{q} - \slashed{l}_2\right) A_{\nu\beta}\right]\right\}_{k = p_1} &\to \left(g_T^{\beta\lambda} \epsilon^{\alpha p n S_T} - g_T^{\alpha\lambda} \epsilon^{\beta p n S_T}\right)\gamma_\lambda \gamma_\beta \frac{q^+ \gamma^-}{(q - l_2)^2}\gamma_\nu\\
& = \left(2 \epsilon^{\alpha P n S_T} - \epsilon^{\beta P n S_T} \gamma^\alpha \gamma_\beta \right)\frac{q^+ \gamma^-}{(q-l_2)^2}\gamma_\nu\,.
\end{split}
\ee
It is clear that the divergent terms in the fourth and sixth lines cancel up to a piece
\be
-\frac{2 p_2^+}{p_1^+} \epsilon^{\alpha P n S_T}\frac{q^+ \gamma^-}{(q - l_2)^2}\gamma_\nu\,,
\ee
which is exactly what we need to cancel the divergence in the second line in~\eqref{eq:divg2}.

\section{Useful integrals}
\label{sec:integral}

Here we list the integrals over $\phi_2$ (azimuthal angle of the parton with momentum $l_2$) that we have encountered in the calculation of the hard coefficients:
 \be
\begin{split}
& \int_0^{2\pi} d\phi_2 \frac{1}{a+b\cos(\phi_1-\phi_2)} = \frac{2\pi {\rm sgn}(a)}{\sqrt{a^2-b^2}} \,, \\
&  \int_0^{2\pi} d\phi_2 \frac{1}{(a+b\cos(\phi_1-\phi_2))^2} = \frac{2\pi |a|}{(a^2-b^2)^{3/2}}
 \,, \\
& \int_0^{2\pi} d\phi_2 \frac{\sin (\phi_2-\Phi_S)}{a+b\cos(\phi_1-\phi_2)} =\frac{2\pi}{b}\sin (\phi_1-\Phi_S)\left(1-\frac{|a|}{\sqrt{a^2-b^2}}\right)\,, \\ 
& \int_0^{2\pi} d\phi_2 \frac{\sin (\phi_2-\Phi_S)}{(a+b\cos(\phi_1-\phi_2))^2} =-2\pi b \sin (\phi_1-\Phi_S)\frac{{\rm sgn}(a)}{(a^2-b^2)^{3/2}}\,, \\ 
& \int_0^{2\pi} d\phi_2 \frac{\cos (\phi_2-\Phi_S)}{a+b\cos(\phi_1-\phi_2)} =\frac{2\pi}{b}\cos (\phi_1-\Phi_S)\left(1-\frac{|a|}{\sqrt{a^2-b^2}}\right)\,, \\ 
& \int_0^{2\pi} d\phi_2 \frac{\cos (\phi_2-\Phi_S)}{(a+b\cos(\phi_1-\phi_2))^2} =-2\pi b \cos (\phi_1-\Phi_S)\frac{{\rm sgn}(a)}{(a^2-b^2)^{3/2}}\,.
\end{split}
\ee

\bibliographystyle{h-physrev}
\bibliography{references}

\end{document}